\documentclass[a4paper,12pt]{article}

\usepackage[utf8]{inputenc}
\usepackage{amssymb,amsmath}
\usepackage[english]{babel}

\newcommand{\MSbar}{\overline{\rm MS}}
\newcommand{\asbar}{\overline{a}_s}

\usepackage[a4paper, total={7in, 9in}]{geometry}
\usepackage{wrapfig}
\usepackage{epstopdf}
\usepackage{graphicx}
\usepackage{multirow}
\usepackage{mathtools}
\usepackage{tabularx}
\usepackage{makecell}
\usepackage{caption}
\usepackage[colorlinks,citecolor=blue,urlcolor=blue,bookmarks=false
]{hyperref}
\usepackage{verbatim}
\usepackage{authblk}
\usepackage{caption}
\RequirePackage{flushend}
\RequirePackage[numbers,sort&compress]{natbib}
\usepackage[binary-units]{siunitx}
\usepackage{tikz}
\usetikzlibrary{arrows,shadows}
\usepackage{setspace}   
\usepackage{ulem}         

\pdfoutput=1

\begin{document}
\pagenumbering{gobble}
\date{\today}

\title{\vspace{-1.5cm}\begin{flushright}
\small{INR-TH-2021-015}
\end{flushright} \vspace{1.5cm} \textbf{The $\MSbar$-scheme 
$\alpha_s^5$ QCD contributions 
to the Adler function and Bjorken polarized sum rule in the Crewther-type two-fold $\{\beta\}$-expanded representation
}}

\author[1]{ I.~O.~Goriachuk\footnote{io.gorjachuk@physics.msu.ru}}
\author[2]{ A.~L.~Kataev\footnote{kataev@ms2.inr.ac.ru}}
\author[2, 3, 4]{ V.~S.~Molokoedov\footnote{viktor\_molokoedov@mail.ru}}

\affil[1]{{\footnotesize Moscow State University,
Faculty of Physics, Department of Theoretical Physics, 119991, Moscow, Russia}}
\affil[2]{{\footnotesize Institute for Nuclear Research of the Russian Academy of Science,
 117312, Moscow, Russia}}
\affil[3]{{\footnotesize Research Computing Center, Moscow State University, 119991, Moscow, Russia}}
\affil[4]{{\footnotesize Moscow Institute of Physics and Technology, 141700, Dolgoprudny, Moscow Region, Russia}}

\maketitle

\begin{abstract}
We consider the two-fold expansion in powers of the conformal anomaly and of the strong coupling $\alpha_s$ for the non-singlet
contributions to Adler $D$-function and Bjorken polarized sum rule calculated previously 
in the $\MSbar$-scheme at the four-loop level.
This representation provides relations between definite terms of different loop orders appearing within the $\{\beta\}$-expansion of these quantities. Supposing the validity of this two-fold representation at the five-loop order and using these relations, we obtain some $\mathcal{O}(\alpha_s^5)$ corrections to the $D$-function, to the
$R$-ratio of $e^+e^-$-annihilation into hadrons and to Bjorken polarized sum rule.
These corrections are presented both analytically in the case of the generic simple gauge
group and numerically for the $SU(3)$ color group. The arguments in the favor of validity of the two-fold representation are given at least at the four-loop level. Within the $\{\beta\}$-expansion procedure 
the analytical Riemann $\zeta_4$-contributions to the five-loop expressions for the Adler function and Bjorken polarized sum rule are also fixed for the case of the generic simple gauge group.
\end{abstract}

\newpage

\section{Introduction}
\pagenumbering{arabic}
\pagestyle{plain}
\setcounter{page}{1}

One of the most informative processes of Particle Physics is the $e^+e^-$  
annihilation into hadrons.
It allows to  probe  the structure of hadronic resonances, which is necessary for testing
 various  strong interaction models including QCD.
Moreover, the investigation of the high-energy behavior of $e^+e^-\rightarrow \gamma^*, Z^* \rightarrow \textit{hadrons}$  enables to study predictions of the Standard 
Model and its possible extensions on the high energies electron-positron  colliders existing at present and planned in future (for plans of their construction see e.g. \cite{Benedikt:2020ejr, 
Freitas:2021oiq}). The important characteristics of the process $e^+e^-\rightarrow \gamma^*\rightarrow {hadrons}$ is the $R$-ratio, which is proportional to its total cross-section.
The function $R\left(s\right)$ is defined in the Minkowski space-time region  and depends on the Mandelstam variable
$s = \left(P_{e^+} + P_{e^-}\right)^2$.
The $R$-ratio is related to the Adler $D$-function and to the hadronic vacuum polarization function $\Pi(Q^2)$, defined in the Euclidean space, by the following way:
\begin{equation}
\label{RtoDint-rel}
D(Q^2) =-Q^2\frac{d\Pi(Q^2)}{dQ^2}= Q^2 \int\limits_0^\infty \frac{R\left(s\right)}{\left(s+Q^2\right)^2} \;d s.
\end{equation}

The QCD expression for the $D$-function in particular represents a special theoretical interest, because the 
renormalization group (RG) equation, determined in the Euclidean domain for this quantity, does not include anomalous dimensions in the limit of massless quarks. 
This means its momentum dependence in perturbation theory (PT) can be totally absorbed into the running of the parameter $\overline{a}_s\equiv\overline{a}_s(Q^2)=\overline{\alpha}_s(Q^2)/\pi$, which we will consider in the modified minimal-subtraction 
$\MSbar$-scheme. The dependence of the  renormalized strong coupling $a_s\equiv a_s(\mu^2)=\alpha_s(\mu^2)/\pi$ on scale $\mu^2$ is described by the solution of the standard RG equation
\begin{equation}
\label{beta-exp}
\beta^{(N)}(a_s ) = \mu^2 \frac{\partial a_s}{\partial \mu^2} =
- \sum\limits_{i = 0}^{N-1} \beta_{i} a^{\;i+2}_s, 
\end{equation}
where $N\geq 1$ is the order of the PT approximation of the $\beta$-function, which 
equals the number of loops involved in the charge renormalization.

The continuing  advances in developing the analytical calculation technique and
its  computer algorithmization  allowed  to obtain the analytic expression for
coefficients of $\beta(a_s)$-function up to the 5-th order of PT in the $\MSbar$-scheme. Indeed, the corresponding  coefficient $\beta_4$ was independently computed by three groups of authors \cite{Baikov:2016tgj, Herzog:2017ohr, Luthe:2017ttc}.

Nowadays, the analytical expression for $D$-function is known at the 4-loop level.  The $\overline{a}^{\;2}_s$-correction in the $\MSbar$-scheme was first independently obtained in  analytical    \cite{Chetyrkin:1979bj}  
and in numerical \cite{Dine:1979qh} form. These results were confirmed  by 
analytical calculation \cite{Celmaster:1979xr}. The correct analytical expression for the  next-to-next-to-leading order (${\rm{NNLO}}$)  $\overline{a}^{\;3}_s$-correction was first obtained in \cite{Gorishnii:1990vf} and was afterwards verified by a non-independent computation in \cite{Surguladze:1990tg}. In this regard, 
the results of totally self-sufficient analytical calculations \cite{Chetyrkin:1996ez}, performed in the arbitrary covariant gauge with help of the completely different technique, were really very important and confirmed those received in \cite{Gorishnii:1990vf}.

The analytical calculations of the separate contributions to the total 4-th order 
correction to $D$-function had been continued more than 15 years. The explicit results for the  
renormalon-type terms were obtained  in \cite{Broadhurst:1993ru}
from the related  direct 5-loop  QED computations of \cite{Broadhurst:1992si}. 
The $\overline{a}^{\;4}_s$-contribution proportional to $n_f^2$ was calculated in \cite{Baikov:2001aa}, while the complete 4-th order results 
for the Adler function are now known from the advanced analytical computations fulfilled in 
\cite{Baikov:2008jh} for the case of $SU(3)$ color group and in \cite{Baikov:2010je, Baikov:2012zn} for the generic simple gauge group.
Note that the independent analytical confirmation of these results was done more than five years later in \cite{Herzog:2017dtz}. 

The concrete self-consistent phenomenological applications of the
RG-improved 4-order PT terms for the $D$-function (for $R$-ratio and Bjorken polarized sum rule as well) 
presume knowledge of the 4-loop $\beta$-function defining the running effect of the strong coupling 
via the solution 
of Eq.(\ref{beta-exp}). At this level the $\MSbar$-scheme function $\beta(a_s)$ was analytically evaluated in \cite{vanRitbergen:1997va} and confirmed  later on in \cite{Czakon:2004bu}. To use in the analysis of the existing $e^+e^-\rightarrow\gamma^*\rightarrow{hadrons}$ data the 5-loop term of the  $\beta(a_s)$-function \cite{Baikov:2016tgj, Herzog:2017ohr, Luthe:2017ttc}, it is preferable to know at least partially the magnitudes of the individual $\overline{a}^{\;5}_s$ contributions to the mentioned quantities or estimates of their $\mathcal{O}(\overline{a}^{\;5}_s)$ total PT corrections (see e.g. \cite{Baikov:2008jh, Boito:2018rwt}).
  
In this work we examine another problem, namely the fixation of the definite $\MSbar$-scheme $\overline{a}^{\;5}_s$ contributions to the still totally analytically unknown five-loop corrections to the considered renorm-group quantities. For this purpose we follow the approach of works \cite{Kataev:2010dm, Kataev:2010du, Cvetic:2016rot}, consisting in the representation of the 5-th order expressions for the $e^+e^-$-annihilation $D$-function and $R$-ratio (and by the same analogy for the Bjorken polarized sum rule), written for the case of the generic simple gauge group, in  the $\{\beta\}$-expanded form introduced in Ref.\cite{Mikhailov:2004iq}. As a result, we fix 8 out of 12 possible $\{\beta\}$-dependent five-loop terms and define their Lie group structure. Moreover, we demonstrate that out of 4 remaining undetermined terms, 3 will contain the Riemann $\zeta_4$-contributions. We also find 
 their exact group structure. The arguments in favor of the validity of these values, coming from the definite cancellations due to the scale and conformal symmetries, are given.

\section{The one-fold generalization of the Crewther relation}

Consider first the Adler $D$-function. In the massless limit this theoretical quantity is decomposed into a sum of the flavor non-singlet (NS) and singlet (SI) components:
\begin{equation}
\label{DAdler}
D^{(M)}(\overline{a}_s) = d_R \left(\sum_f {Q_f}^2\right) D^{(M)}_{NS} (\overline{a}_s) +
d_R \left(\sum_f Q_f\right)^2 D^{(M\geq 3)}_{SI} (\overline{a}_s),
\end{equation}
where $Q_f$ are the charges of quarks  with flavor $f$,  $d_R$ is the dimension of the quark representation of the considered generic simple gauge group. In our study we are primarily interested in the case of the   $SU(N_c)$ gauge group, where $d_R=N_c$, and its particular case of the $SU(3)$ color group, relevant for physical QCD. The number $M$ stands for the order of PT being considered. 
For the non-singlet case starting from $M\geq 2$ it is related to $N$, defined below  Eq.(\ref{beta-exp}), as $M=N+1$.
The singlet in flavor contribution to $D$-function is appearing first at the third order of 
PT \cite{Gorishnii:1990vf}.

Another RG quantity interesting for our investigation is
the characteristic of the pure Euclidean process of deep inelastic scattering of the polarized leptons on nucleons, namely the Bjorken polarized sum rule, which is defined as:
\begin{equation}
\label{BJPSR}
S^{(M)}_{Bjp}(\overline{a}_s) = \int\limits_0^1 \bigg(g_{1, (M)}^{(lp)}(x,Q^2)-g_{1, (M)}^{(ln)}(x,Q^2)\bigg) dx =\frac{1}{6}\bigg|\frac{g_A}{g_V}\bigg|\bigg( C^{(M)}_{NS}(\overline{a}_s)+ C^{(M\geq 4)}_{SI}(\overline{a}_s)\bigg),
\end{equation}
where $g_{1, (M)}^{(lp)}(x,Q^2)$ and $g_{1, (M)}^{(ln)}(x,Q^2)$ are the structure functions of these deep-inelastic processes, $g_A$ and $g_V$ are the axial and vector neutron $\beta$-decay constants. In Eq.(\ref{BJPSR}) the dependence of $S_{Bjp}$ on $Q^2=-q^2$ at large $Q^2$ is absorbed into the coupling $\overline{a}_s$.

In the case of the generic color gauge group the analytical expressions for the second, third and fourth PT corrections to $C_{NS}(\overline{a}_s)$-function were analytically computed in the $\MSbar$-scheme in Refs.\cite{Gorishnii:1985xm}, \cite{Larin:1991tj}, \cite{Baikov:2010je}  respectively. It should be also reminded that in contrast to the Adler function, the SI contribution to $S_{Bjp}(\overline{a}_s)$ manifests itself starting  from the 4-th order of PT \cite{Larin:2013yba}\footnote{However, the guessed expression of the $\overline{a}^{\;4}_s$ contribution to $C^{(M=4)}_{SI}$ in \cite{Larin:2013yba} does not agree with the analytical result for the same term, directly evaluated in \cite{Baikov:2015tea}. The reason for this discrepancy is still the opened question.}.  

Theoretically interesting feature of 
the Adler function and Bjorken polarized sum rule 
is that their Born approximations are related through  consideration of the axial-vector-vector (AVV) triangle diagram.  In the massless case respecting the conformal symmetry this relation 
was derived by Crewther in \cite{Crewther:1972kn} (see  \cite{Adler:1972msp} for reanalysis and  \cite{Crewther:2014kha} for review).
 The PT generalization 
of the Crewther relation was discovered in \cite{Broadhurst:1993ru} at $M=3$ (or $N=2$) and reads:
\begin{equation} 
\label{BK}
D^{(M)}_{NS} (\overline{a}_s)C^{(M)}_{NS}(\overline{a}_s)=
1+\bigg(\frac{\beta^{(M-1)}(\overline{a}_s)}{\overline{a}_s}\bigg)K^{(M-1)}(\overline{a}_s)+\mathcal{O}(\overline{a}^{\;M+1}_s).
\end{equation} 

Here $\beta^{(2)}(\overline{a}_s)$ is the two-loop approximation of the $\beta$-function (\ref{beta-exp})
of the non-abelian strong interaction theory with a simple compact Lie group. At $M=3$ the term $K^{(2)}(\overline{a}_s)=K_1\overline{a}_s +K_2\overline{a}^{\;2}_s$ is the second-degree polynomial with coefficient $K_1$, containing only the quadratic Casimir operator $C_F$
of the fundamental representation of the gauge group,
and with coefficient $K_2$ that includes three monomials $C_F^2$, $C_FC_A$ and $C_FT_Fn_f$, where $C_A$ is the quadratic Casimir 
operator of the adjoint representation, $T_F$ is the Dynkin index and $n_f$ is the number of quark flavors. 
There was made a guess in \cite{Broadhurst:1993ru}  that the factorization property of the $\beta(\overline{a}_s)$-function in the generalized Crewther relation may be also true at $M>3$ as well. Further on, we will call
the form of Eq.(\ref{BK})  the one-fold  generalization of the Crewther relation.

It follows from consideration of the PT expression for the AVV diagram with three possible form-factors appearing in the special kinematics $(pq)=0$ that the mutual cancellation of the $C_F^M\asbar^M$-factors 
in the product of the Adler function and Bjorken polarized sum rule (see Eq.(\ref{BK})) in all orders of PT is the consequence of the Adler-Bardeen theorem on the non-renormalizability of axial current \cite{Gabadadze:1995ei}\footnote{This 
fact also follows from studies of the perturbative quenched QED \cite{Kataev:2013vua}, previously called ``the finite QED program'' \cite{Johnson:1967pk} and respecting the conformal symmetry.}.
Moreover, it was shown there that 
one of these three form-factors stays 
non-renormalizable, while the remaining two are renormalized and are proportional to the conformal anomaly term $\beta^{(N)}(\asbar)/\asbar$. These
two renormalized form-factors are related to the 
 one-fold generalization of the Crewther relation and 
may be expressed through the factorized $\beta$-function in all orders of PT starting from the second one.  The direct analytical calculations \cite{Mondejar:2012sz} of all 3-loop corrections to 
6 possible form-factors of the AVV Green function in the arbitrary kinematics detect the appearance in the $\overline{a}^{\;2}_s$-contributions to the certain form-factors of the terms, 
 proportional to the first coefficient $\beta_0$ of the QCD $\beta$-function. 
Thus, the aforesaid means that the conjecture concerning the factorization of the conformal anomaly term in the one-fold generalization of the Crewther relation in higher orders of PT \cite{Broadhurst:1993ru, Gabadadze:1995ei} {\it may be} indeed true. The consequent more general theoretical $x$-space analysis of Ref.\cite{Crewther:1997ux} indicates this statement  
ought be changed to the assertion {\it the $\beta(\overline{a}_s)$-function should be factorized}  in all orders of PT. The studies, conducted in Ref.\cite{Braun:2003rp}, lead to the same conclusion.

As was shown in \cite{Garkusha:2018mua}, if this inference is really true in the gauge-invariant $\MSbar$-scheme then it will be valid in the gauge-dependent ${\rm{MOM}}$-like schemes in the Landau gauge in all orders as well\footnote{For the most recent analytical 5-loop calculations of the definite anomalous dimensions in the $\MSbar$-scheme in the Landau gauge see \cite{Chetyrkin:2017bjc}.}.
 
The important explicit 4-th order analytical computations of the PT corrections to
$D^{(4)}_{NS}(\asbar)$ and 
$C^{(4)}_{NS}(\asbar)$-functions, carried out in the $\MSbar$-scheme for the case of a generic simple gauge group in Ref.\cite{Baikov:2010je}, allowed the authors of this work to verify directly and thereby confirm the conjecture
\cite{Broadhurst:1993ru} on the validity of  Eq.(\ref{BK}) at $M=4$ with the factorizable three-loop function $\beta^{(3)}(\asbar)$, which was evaluated in \cite{Tarasov:1980au, Larin:1993tp}.
At $M=4$ the term $K^{(3)}(\asbar)=K_1\asbar+K_2\asbar^{\;2}+K_3\asbar^{\;3}$ is the cubic polynomial in $\asbar$. Its $\asbar^{\;3}$-coefficient $K_3$ was obtained in Ref.\cite{Baikov:2010je}. It 
contains 6 color monomials, namely $C_F^3$, $C_F^2C_A$, $C_FC_A^2$,
$C_F^2T_Fn_f$, $C_FC_AT_Fn_f$, $C_FT^2_Fn^2_f$. One should emphasize that coefficients $K_2$ and $K_3$ contain $n_f$-dependent terms. Now let us move on to the representation of the generalized Crewther relation
where at least in the 4-th order of PT the total flavor dependence of its r.h.s will be contained in coefficients of $\beta$-function only.

\section{The two-fold generalization of the Crewther relation }

At the next stage, using the 3-rd order analytical
PT expressions for $D^{(3)}_{NS}$ and $C^{(3)}_{NS}$-functions, it was found in \cite{Kataev:2010dm} that 
in the case of a generic simple gauge group the generalized Crewther relation  at $2\leq M\leq 3$
can be rewritten in the following two-fold representation:  
\begin{equation} 
\label{BK-two-fold}
D^{(M)}_{NS} (\overline{a}_s)C^{(M)}_{NS}(\overline{a}_s)=
1+\sum\limits_{n=1}^{M-1}\bigg(\frac{\beta^{(M-n)}(\overline{a}_s)}{\overline{a}_s}\bigg)^nP^{(M-n)}_n(\overline{a}_s)+\mathcal{O}(\overline{a}^{\;M+1}_s).
\end{equation} 

The validity of the analogous expression at $M=4$ was assumed in \cite{Kataev:2010dm} even prior the
appearance of the analytical results for $\overline{a}^{\;4}_s$-corrections to $D^{(4)}_{NS}$ and $C^{(4)}_{NS}$-functions \cite{Baikov:2010je}. The explicit check of this guess was demonstrated a bit later in \cite{Kataev:2010du}, where the results of Ref.\cite{Baikov:2010je} had already been used. 

At the 4-loop level the polynomials $P^{(r)}_n(\overline{a}_s)$ in Eq.(\ref{BK-two-fold}) read:
\begin{equation}
\label{polynomial}
P^{(r)}_n(\overline{a}_s)=\sum\limits_{k=1}^r P^{(r)}_{n, k} \;\overline{a}^{\;k}_s=\sum_{p = 1}^{4-n} \overline{a}^{\;p}_s \sum_{k=1}^p P_n^{(r)}[k,p-k] C_F^k C_A^{p-k},  
\end{equation}
where $r=M-n$ and for the considered case (when $M=4$)
$r=4-n$ correspondingly. Coefficients $P^{(r)}_{n, k}$ can be unambiguously defined at least in the 4-th order of PT \cite{Kataev:2010du}. An important point here is that at this level of PT all dependence on $n_f$ in r.h.s. of Eq.(\ref{BK-two-fold}) is held in the coefficients of the RG $\beta$-function. Thus, in contrast to the coefficients of the polynomial $K^{(M-1)}(\overline{a}_s)$ in Eq.(\ref{BK}), the terms of $P^{(r)}_n(\overline{a}_s)$ in Eq.(\ref{BK-two-fold}) are independent on the number of quark flavors. 

\section{The two-fold representation for quantities related by the generalized Crewther relation}

\subsection{The case of the Adler function}
\label{SubSecAdler}
 
The double sum expression for the conformal symmetry breaking term in r.h.s. of Eq.(\ref{BK-two-fold}) motivated the authors of the work 
\cite{Cvetic:2016rot} to consider the similar representation for the NS contributions to the Adler function and Bjorken polarized sum rule at least at the analytically available $\mathcal{O}(\overline{a}^{\;4}_s)$ level. According to this paper at $M=4$ the PT expression for the NS Adler function, calculated in the $\MSbar$-scheme for the non-abelian gauge theory with a simple compact Lie group, may be presented as the following two-fold series:
\begin{equation}
\label{D-two-fold}
D^{(M)}_{NS}(\overline{a}_s) = 1 + D^{(M)}_0(\overline{a}_s)+\sum\limits_{n=1}^{M-1}\left(\frac{\beta^{(M-n)}(\overline{a}_s)}{\overline{a}_s}\right)^{n} D^{(M-n)}_{n}(\overline{a}_s)+\mathcal{O}(\overline{a}^{\;M+1}_s),
\end{equation}
where polynomials $D^{(r)}_n(\overline{a}_s)$ in the coupling constant $\overline{a}_s$ are:
\begin{equation}
\label{Drn}
D^{(r)}_n(\overline{a}_s)=\sum\limits_{k=1}^r D^{(r)}_{n,k}\overline{a}^{\;k}_s.
\end{equation}

In a more detailed form Eq.(\ref{D-two-fold}) may be written down as:
\begin{eqnarray}
\label{explicit}
\hspace{-0.7cm}
D^{(M=4)}_{NS}(\overline{a}_s)&=&1 + D^{(4)}_{0,1}\overline{a}_s+\bigg(D^{(4)}_{0,2}-\beta_0 D^{(3)}_{1,1}\bigg)\overline{a}^{\;2}_s+\bigg(D^{(4)}_{0,3}-\beta_0 D^{(3)}_{1,2}-\beta_1 D^{(3)}_{1,1}+\beta^2_0 D^{(2)}_{2,1}\bigg)\overline{a}^{\;3}_s \\ \nonumber
&+&\bigg(D^{(4)}_{0,4}-\beta_0 D^{(3)}_{1,3}-\beta_1 D^{(3)}_{1,2}-\beta_2 D^{(3)}_{1,1}+\beta^2_0 D^{(2)}_{2,2}+2\beta_0\beta_1 D^{(2)}_{2,1}-\beta^3_0 D^{(1)}_{3,1}\bigg)\overline{a}^{\;4}_s+\mathcal{O}(\overline{a}^{\;5}_s),
\end{eqnarray}
where at the fixed number of $n$ and $k$,  $D^{(r)}_{n,k}\equiv D^{(r+1)}_{n,k}\equiv D^{(r+2)}_{n,k}\equiv \dots$, e.g. the terms $D^{(2)}_{1,2}\equiv D^{(3)}_{1,2}$.

The coefficients $D^{(r)}_{n,k}$ including in Eq.(\ref{explicit}) are determined by an unambiguous way
as solutions of a system of linear equations, analogous to those presented in \cite{Kataev:2010du}. Herewith, the full dependence on $n_f$ (except for the light-by-light scattering effects -- see explanations below) is absorbed into the coefficients of $\beta$-function and their combinations (\ref{explicit}). For the first time at the four-loop level the values of $D^{(r)}_{n,k}$ were obtained in the work \cite{Cvetic:2016rot}. Note also that the two-fold representation (\ref{D-two-fold}) and its counterpart for the Bjorken polarized sum rule to be considered below in Subsec. \ref{Bjsub} are a sufficient condition for the validity of the generalized Crewther relation in the form of Eq.(\ref{BK-two-fold}).

Starting from the four-loop level the coefficients of the NS Adler function in the case of the generic simple gauge group contain contributions of the light-by-light scattering type, proportional to the group structures $d^{abcd}_Fd^{abcd}_A/d_R$ and $d^{abcd}_Fd^{abcd}_Fn_f/d_R$ \cite{Baikov:2010je}.
Here $d^{abcd}_F={\rm{Tr}}(T^aT^{\mathop{\{}b}T^cT^{d\mathop{\}}})/6$ and $d^{abcd}_A={\rm{Tr}}(C^aC^{\mathop{\{}b}C^cC^{d\mathop{\}}})/6$, where
the symbol $\{\dots\}$ stands for the full symmetrization procedure of elements $T^bT^cT^d$ by superscripts $b$, $c$ and $d$ \cite{vanRitbergen:1997va}; $T^a$ are the generators of the representation of fermions, $(C^a)_{bc}=-if^{abc}$ are the generators of the adjoint representation with the antisymmetric structure constants $f^{abc}$ of the Lie algebra: $[T^a,T^b]=if^{abc}T^c$. In the case of the $SU(N_c)$  color gauge group the completeness relation for the generators in the defining representation of its Lie algebra leads to the following expressions for the aforementioned contractions (together with $d^{abcd}_Ad^{abcd}_A/d_R$):
\begin{subequations}
\begin{gather}
\label{color-structures-1}
\frac{d^{abcd}_Fd^{abcd}_F}{d_R}=\frac{(N^4_c-6N^2_c+18)(N^2_c-1)}{96N^3_c}, ~~~~ \frac{d^{abcd}_Fd^{abcd}_A}{d_R}=\frac{(N^2_c+6)(N^2_c-1)}{48}, \\ \frac{d^{abcd}_Ad^{abcd}_A}{d_R}=\frac{N_c(N^2_c+36)(N^2_c-1)}{24}.
\end{gather}
\end{subequations}

Although the term $d_F^{abcd}d_F^{abcd}n_f \overline{a}^{\;4}_s/d_R$ in $D^{(4)}_{NS}(\overline{a}_s)$  is proportional
to the number of flavors $n_f$, which formally enters the $\beta_0$-coefficient, we will not include it
into $D^{(3)}_{1,3}$-coefficient in Eq.(\ref{explicit}), since such embedding will not be supported by the QED limit \cite{Cvetic:2016rot}. Indeed, in the QED limit of the QCD-like theory with the $SU(N_c)$ group $d_F^{abcd}d^{abcd}_F/d_R=1$ and $n_f=N$, where $N$ is the number of the charged leptons (structures with $d^{abcd}_A$ are nullified). This term arises from the five-loop Feynman diagram with light-by-light scattering internal subgraphs (fermion loop with four  photon propagators coming out of it), contributing to the photon vacuum polarization function. However, in QED the sum of these subgraphs are convergent and does
not give extra $\beta_0$-dependent (or $N$-dependent) contribution to the coefficient $D^{(3)}_{1,3}$ \cite{Cvetic:2016rot}. Therefore, to get a smooth transition from the case of $U(1)$ to $SU(N_c)$ gauge group,  these light-by-light scattering terms should be included into the $\beta$-independent coefficient $D^{(4)}_{0,4}$ at the four-loop level.

The foregoing enables to rewrite Eq.(\ref{Drn}) at the four-loop level in the form of the
explicit decomposition by the Casimir operator:
\begin{eqnarray}
\label{Dn_double-sum}
D^{(r)}_n(\overline{a}_s) &=& \sum_{p = 1}^{4-n} \overline{a}^{\;p}_s \sum_{k=1}^p D_n^{(r)}[k,p-k] C_F^k C_A^{p-k}  \nonumber \\
&+& \overline{a}^{\;4}_s \delta_{n0}  \bigg(D_0^{(4)}[F,A] \frac{d_F^{abcd}d_A^{abcd}}{d_R} +
D_0^{(4)}[F,F] \frac{d_F^{abcd}d_F^{abcd}}{d_R}n_f\bigg),
\end{eqnarray}
where terms with the Kronecker symbol correspond to the light-by-light scattering effects discussed above.
Up to the 4-th order of PT the coefficients $D_n^{(r)}[k, p-k]$ are known analytically in terms of
rational numbers and the odd Riemann $\zeta$-functions, namely $\zeta_3$, $\zeta_5$, $\zeta_7$, $\zeta^2_3$ \cite{Cvetic:2016rot} (see details below). Here as before at the fixed number of $n$, $k$ and $p$, the coefficients $D^{(r)}_{n}[k, p-k]\equiv D^{(r+1)}_{n}[k, p-k]\equiv D^{(r+2)}_{n}[k, p-k]\equiv \dots$,  e.g.  $D^{(2)}_1[2,0]=D^{(3)}_1[2,0]=D^{(4)}_1[2,0]$.

It turns out that the two-fold representation of the NS Adler function, whose detailed form at the $\mathcal{O}(\overline{a}^{\;4}_s)$ level is given in Eq.(\ref{explicit}), is in full agreement with its $\{\beta\}$-expansion structure proposed in Ref.\cite{Mikhailov:2004iq} 12 years earlier than the representation (\ref{D-two-fold}). Indeed, accordingly to this work the coefficients $d_M$ $(1\leq M\leq 4)$ of the NS Adler function
\begin{equation}
\label{simD}
D^{(M=5)}_{NS}(\overline{a}_s)=1+d_1\overline{a}_s+d_2\overline{a}^{\;2}_s+d_3\overline{a}^{\;3}_s+d_4\overline{a}^{\;4}_s+d_5\overline{a}^{\;5}_s + \mathcal{O}(\overline{a}^{\;6}_s)
\end{equation}
may be decomposed into the coefficients of the $\beta(\overline{a}_s)$-function and their definite combinations and have the following form:
\begin{subequations}
\begin{gather}
\label{d1}
d_1 = d_1[0], \\ 
\label{d2}
d_2 = \beta_0 d_2[1] + d_2[0],  \\ 
\label{d3}
d_3 = \beta_0^2 d_3[2] + \beta_1 d_3[0,1] + \beta_0 d_3[1] + d_3[0],  \\ 
\label{d4}
d_4 = \beta_0^3 d_4[3] + \beta_1 \beta_0 d_4[1,1] + \beta_2 d_4[0,0,1] + \beta_0^2 d_4[2] + \beta_1  d_4[0,1] + \beta_0 d_4[1] + d_4[0]. 
\end{gather}

Relations (\ref{d1}-\ref{d4}) are in full compliance with the results of application of the two-fold representation (\ref{explicit}).

Using the two-fold form (\ref{explicit}), the authors of work \cite{Cvetic:2016rot} obtained all $\{\beta\}$-expanded terms in $d_2$, $d_3$ and $d_4$ coefficients in the $\MSbar$-scheme. These results are
presented in Tables \ref{T-d1-3} and \ref{T-d4}.

\begin{table}[h!]
\renewcommand{\tabcolsep}{0.6cm} 
\renewcommand{\arraystretch}{1.7}
\centering
\begin{tabular}{|c|c|c|}
\hline
Coefficients              & Group structures &                                                  Numbers    \\ \hline
$d_1[0]$                  & $C_F$            & $\frac{3}{4}$                                         \\ \hline

 & $C^2_F$ & $-\frac{3}{32}$                                       \\ \cline{2-3} 
 
\multirow{-2}{*}{$d_2[0]$}& $C_FC_A$         & $\frac{1}{16}$                                        \\ \hline
$d_2[1]$                  & $C_F$            & $\frac{33}{8}-3\zeta_3$                               \\ \hline
 & $C^3_F$          & $-\frac{69}{128}$                                     \\ \cline{2-3} 
 & $C^2_FC_A$       & $-\frac{101}{256}+\frac{33}{16}\zeta_3$               \\ \cline{2-3} 

\multirow{-3}{*}{$d_3[0]$}  & $C_FC^2_A$       & $-\frac{53}{192}-\frac{33}{16}\zeta_3$                \\ \hline
                                         
\multirow{2}{*}{$d_3[1]$} & $C^2_F$          & $-\frac{111}{64}-12\zeta_3+15\zeta_5$                 \\ \cline{2-3} 
                          & $C_FC_A$         & $\frac{83}{32}+\frac{5}{4}\zeta_3-\frac{5}{2}\zeta_5$ \\ \hline
$d_3[0,1]$                & $C_F$            & $\frac{33}{8}-3\zeta_3$                               \\ \hline
$d_3[2]$                  & $C_F$            & $\frac{151}{6}-19\zeta_3$                             \\ \hline
\end{tabular}
\captionsetup{justification=centering}
\caption{\label{T-d1-3} All terms included in the $\{\beta\}$-expansion of coefficients $d_1$, $d_2$ and $d_3$.}
\end{table}

\begin{table}[h!]
\renewcommand{\tabcolsep}{0.6cm} 
\renewcommand{\arraystretch}{1.7}
\centering
\begin{tabular}{|c|c|c|}
\hline
Coefficients                                                              & Group structures                      &                                                                                                                            Numbers \\ \hline
 & $C^4_F$                               & $\frac{4157}{2048}+\frac{3}{8}\zeta_3$                                                                                      \\ \cline{2-3} 
 & $C^3_FC_A$                            & $-\frac{3509}{1536}-\frac{73}{128}\zeta_3-\frac{165}{32}\zeta_5$                                                            \\ \cline{2-3} 
 & $C^2_FC^2_A$                          & $\frac{9181}{4608}+\frac{299}{128}\zeta_3+\frac{165}{64}\zeta_5$                                                            \\ \cline{2-3} 
 & $C_FC^3_A$                            & $-\frac{30863}{36864}-\frac{147}{128}\zeta_3+\frac{165}{64}\zeta_5$                                                         \\ \cline{2-3} 
 & $\frac{d^{abcd}_Fd^{abcd}_A}{d_R}$    & $\frac{3}{16}-\frac{1}{4}\zeta_3-\frac{5}{4}\zeta_5$                                                                        \\ \cline{2-3} 
\multirow{-6}{*}{$d_4[0]$} & $\frac{d^{abcd}_Fd^{abcd}_F}{d_R}n_f$ & $-\frac{13}{16}-\zeta_3+\frac{5}{2}\zeta_5$                                                                                 \\ \hline
                                                                          & $C^3_F$                               & $-\frac{785}{128}-\frac{9}{16}\zeta_3+ \frac{165}{2}\zeta_5-\frac{315}{4}\zeta_7$ \\ \cline{2-3} 
                                                                          & $C^2_FC_A$                            & $-\frac{3737}{144}+\frac{3433}{64}\zeta_3-\frac{99}{4}\zeta^2_3-\frac{615}{16}\zeta_5+\frac{315}{8}\zeta_7$                 \\ \cline{2-3} 
\multirow{-3}{*}{$d_4[1]$}                                                & $C_FC^2_A$                            & $-\frac{2695}{384}-\frac{1987}{64}\zeta_3+\frac{99}{4}\zeta^2_3+\frac{175}{32}\zeta_5-\frac{105}{16}\zeta_7$                \\ \hline
                                                                          & $C^2_F$                               & $-\frac{111}{64}-12\zeta_3+15\zeta_5$                                                                                       \\ \cline{2-3} 
\multirow{-2}{*}{$d_4[0,1]$}                                              & $C_FC_A$                              & $\frac{83}{32}+\frac{5}{4}\zeta_3-\frac{5}{2}\zeta_5$                                                                       \\ \hline
                                                                          & $C^2_F$                               & $-\frac{4159}{384}-\frac{2997}{16}\zeta_3+27\zeta^2_3+\frac{375}{2}\zeta_5$                                                 \\ \cline{2-3} 
\multirow{-2}{*}{$d_4[2]$}                                                & $C_FC_A$                              & $\frac{14615}{256}+\frac{39}{16}\zeta_3-\frac{9}{2}\zeta^2_3-\frac{185}{4}\zeta_5$                                          \\ \hline
$d_4[0,0,1]$                                                              & $C_F$                                 & $\frac{33}{8}-3\zeta_3$                                                                                                     \\ \hline
$d_4[1,1]$                                                                & $C_F$                                 & $\frac{151}{3}-38\zeta_3$                                                                                                   \\ \hline
$d_4[3]$                                                                  & $C_F$                                 & $\frac{6131}{36}-\frac{203}{2}\zeta_3-45\zeta_5$                                                                            \\ \hline
\end{tabular}
\captionsetup{justification=centering}
\caption{\label{T-d4} All terms included in the $\{\beta\}$-expansion of coefficient $d_4$.}
\end{table}

Note once again that the $\{\beta\}$-expanded terms of the coefficients $d_2$ and $d_3$ are independent on the number of flavors.

As was already discussed above and was mentioned in Appendix of Ref.\cite{Brodsky:2013vpa},
 the term $d_4[0]$ contains $d^{abcd}_Fd^{abcd}_A/d_R$ and $d^{abcd}_Fd^{abcd}_Fn_f/d_R$ contributions of the light-by-light scattering type (see Table \ref{T-d4}) that satisfies the correct transition to the QED limit.

Supposing the counterpart of representation (\ref{D-two-fold}) to be true at the $\mathcal{O}(\overline{a}^{\;5}_s)$ level $(M=5, N=4)$, one can obtain the analogous $\{\beta\}$-expansion for $d_5$-coefficient in (\ref{simD}), which is consistent with the one presented in \cite{Mikhailov:2016feh}:
\begin{eqnarray} 
\label{d5}
d_5 &=& \beta^4_0 d_5[4] + \beta_1\beta^2_0 d_5[2,1] + \beta^3_0 d_5[3] + \beta_2 \beta_0 d_5[1,0,1] + \beta^2_1 d_5[0,2]  
+ \beta_1 \beta_0 d_5[1,1] \nonumber \\
&+& \beta^2_0 d_5[2] + 
 \beta_3 d_5[0,0,0,1] + \beta_2 d_5[0,0,1] + \beta_1 d_5[0,1] + \beta_0 d_5[1] + d_5[0]. 
\end{eqnarray}   
\end{subequations}

Moreover, expanding the right-hand side of the analog of Eq.(\ref{D-two-fold}) at $M=5$ in the series in coupling constant,
one can observe that the different coefficients of the $\beta$-function in various combinations are
multiplied by the same coefficients $D^{(r)}_{n,k}$ for any $n$ (except for $n=0$).
This leads to the definite relations between the coefficients of $\{\beta\}$-expansions (\ref{d2}-\ref{d5}).
One can write them in a tabular-like manner:
\begin{subequations}
\begin{eqnarray}
\label{D_{1,1}}
D_{1,1}^{(1)} &=& -d_2[1] = -d_3[0,1] = -d_4[0,0,1] = -d_5[0,0,0,1],  \\
\label{D_{1,2}}
D_{1,2}^{(2)} &=& -d_3[1] = -d_4[0,1] = -d_5[0,0,1],  \\
\label{D_{1,3}}
D_{1,3}^{(3)} &=& -d_4[1] = -d_5[0,1]; \\
\label{D_{2,1}}
D_{2,1}^{(1)} &=& d_3[2] =  d_4[1,1]/2 = d_5[0,2] =  d_5[1,0,1]/2,  \\
\label{D_{2,2}}
D_{2,2}^{(2)} &=& d_4[2] =  d_5[1,1]/2;  \\
\label{D_{3,1}}
D_{3,1}^{(1)} &=& -d_4[3] = - d_5[2,1]/3. 
\end{eqnarray}
\end{subequations}

In accordance with the aforesaid, for instance, the coefficient $D^{(1)}_{1,1}\equiv D^{(2)}_{1,1}\equiv D^{(3)}_{1,1}$, $D^{(3)}_{1,3}$ \footnote{
It should be noticed that 
there is a misprint in the paper \cite{Cvetic:2016rot} in the formula (9). Instead of the rational $C^3_F$-coefficient $758/128$ in the expression for $D_1(\overline{a}_s)$ should be $785/128$ (see $d_4[1]$ in Table \ref{T-d4}).}$\equiv D^{(4)}_{1,3}$.

Due to the relations (\ref{D_{1,1}}-\ref{D_{3,1}}) it is possible to restore 7 out of 12
coefficients in the $\{\beta\}$-expansion of $d_5$  (\ref{d5}). Another one, namely $d_5[4]$, is known  and corresponds to renormalon contributions whose general formula was obtained in \cite{Broadhurst:1993ru}.
All $\{\beta\}$-expanded coefficients of $d_5$, obtained in this way, are summarized in the Table \ref{T-d5}.

\begin{table}[h!]
\renewcommand{\tabcolsep}{0.6cm} 
\renewcommand{\arraystretch}{1.8}
\centering
\begin{tabular}{|c|c|c|}
\hline \vspace{-0.2cm}
~~~Coefficients~~~                 & ~~~Group structures~~~ &                                                                                                            Numbers \\ \hline
\multirow{3}{*}{$d_5[0,1]$}   & 
$C^3_F$          &  $-\frac{785}{128}-\frac{9}{16}\zeta_3+\frac{165}{2}\zeta_5-\frac{315}{4}\zeta_7$  \\  \cline{2-3}
  & $C^2_FC_A$       & ~~$-\frac{3737}{144}+\frac{3433}{64}\zeta_3-\frac{99}{4}\zeta^2_3-\frac{615}{16}\zeta_5+\frac{315}{8}\zeta_7$~~ \\  \cline{2-3} 
                              & $C_FC^2_A$       & $-\frac{2695}{384}-\frac{1987}{64}\zeta_3+\frac{99}{4}\zeta^2_3+\frac{175}{32}\zeta_5-\frac{105}{16}\zeta_7$       \\   \hline
\multirow{2}{*}{$d_5[0,0,1]$} & $C^2_F$          & $-\frac{111}{64}-12\zeta_3+15\zeta_5$                                                                       \\  \cline{2-3} 
                              & $C_FC_A$         & $\frac{83}{32}+\frac{5}{4}\zeta_3-\frac{5}{2}\zeta_5$                                                       \\   \hline
$d_5[0,0,0,1]$                & $C_F$           & $\frac{33}{8}-3\zeta_3$
\\ \hline                  
\multirow{2}{*}{$d_5[1,1]$}   & $C^2_F$          & $-\frac{4159}{192}-\frac{2997}{8}\zeta_3+54\zeta^2_3+375\zeta_5$                                            \\   \cline{2-3} 
                              & $C_FC_A$         & $\frac{14615}{128}+\frac{39}{8}\zeta_3-9\zeta^2_3-\frac{185}{2}\zeta_5$                                     \\    \hline
$d_5[0,2]$                    & $C_F$            & $\frac{151}{6}-19\zeta_3$                                                                                   \\     \hline
$d_5[1,0,1]$                  & $C_F$            & $\frac{151}{3}-38\zeta_3$                                                                                   \\      \hline
$d_5[2,1]$                    & $C_F$            & $\frac{6131}{12}-\frac{609}{2}\zeta_3-135\zeta_5$                                                           \\      \hline
$d_5[4]$                      & $C_F$            & $\frac{91865}{72}-\frac{4955}{9}\zeta_3-570\zeta_5$                                                         \\      \hline
\end{tabular}
\captionsetup{justification=centering}
\caption{\label{T-d5} All extracted terms included in the $\{\beta\}$-expansion of coefficient $d_5$.}
\end{table}

In analogous way, assuming the validity of the two-fold representation at the six-loop level, it is possible to restore 11 out of 19 terms (including the corresponding renormalon one from \cite{Broadhurst:1993ru}) for the $\{\beta\}$-expanded coefficient $d_6$ of $D$-function  and etc.

In the numerical form for the case of $SU(3)$ group the presented $\{\beta\}$-expanded coefficients read:
\begin{eqnarray}
\label{DNS-beta-concrete}
\nonumber
D^{(M=5)}_{NS}(\overline{a}_s)&=&1+\overline{a}_s+\bigg(0.6918\beta_0+\uuline{0.0833}\bigg)\overline{a}^{\;2}_s +\bigg(3.1035\beta^2_0
+0.6918\beta_1+4.9402\beta_0-\uuline{23.2227}\bigg)\overline{a}^{\;3}_s \\ \nonumber
&+&\bigg(2.1800\beta^3_0+6.2069\beta_0\beta_1
+17.6990\beta^2_0+
0.6918\beta_2+4.9402\beta_1 
-101.928\beta_0 \\ \nonumber
&+&\uuline{81.1571}+0.0802n_f\bigg)\overline{a}^{\;4}_s 
+\bigg(30.7398\beta^4_0+6.5401\beta^2_0\beta_1
+6.2069\beta_0\beta_2+3.1035\beta^2_1 \\ \nonumber
&+& 35.3981\beta_0\beta_1 
+0.6918\beta_3+4.9402\beta_2-101.928\beta_1 \\
&+&\uwave{d_5[3]}\beta^3_0+\uwave{d_5[2]}\beta^2_0+\uwave{d_5[1]}\beta_0+\uwave{d_5[0]}\bigg)\overline{a}^{\;5}_s+\mathcal{O}(\overline{a}^{\;6}_s)
\end{eqnarray}

Thus, 4 wavy terms out of 12 possible ones remain undetermined in $d_5$ (but with the fixed $\zeta_4$-contributions to $d_5[0], d_5[1], d_5[2]$ -- see extra clarifications below).

It is worth emphasizing that the first two solid underlined conformal-invariant $\beta$-independent terms are in agreement with ones, obtained with help of the generalized Brodsky-Lepage-Mackenzie\footnote{The somewhat different extension of the BLM method, called ``sequental extended BLM'' (seBLM),  was proposed in \cite{Mikhailov:2004iq}, where all $\{\beta\}$-expanded terms are absorbed into the coupling-dependent scale.} (BLM \cite{Brodsky:1982gc}) ${\rm{MS}}$-like scale fixing prescription \cite{Grunberg:1991ac}\footnote{
Note that the term $d_2[0]=0.0833$, obtained within the pure QCD, is reproduced in the skeleton approach 
\cite{Cvetic:2006gc} 
and in the minimally extended SUSY QCD with light gluinos \cite{Kataev:2010du, Mikhailov:2004iq, Mikhailov:2016feh, Kataev:2014jba}, whereas the third-order conformally-invariant contribution $d_3[0]=-23.2227$ differs from 
corresponding results $-27.849$ of \cite{Cvetic:2006gc}
and  $-35.8725$ of \cite{Kataev:2010du, Mikhailov:2004iq, Mikhailov:2016feh, Kataev:2014jba}. For comparison with the gluino-extended QCD results see Sec.\ref{Discussion} of this paper.}.
Further on, the numerical expression for the 4-th 
 order conformal invariant term are confirmed by the 
  results of the first realization  \cite{Brodsky:2011ta} of the ideas of the Principle of Maximum Conformality (PMC/BLM) \cite{Brodsky:2011ig}. Like the seBLM prescription, the PMC is a generalization of the BLM procedure to higher orders of PT, enabling to systematically absorb all $\beta$-dependent terms of $d_M$-coefficients in the scale parameter. 
  
These coincidences of the underlined terms with the results of the PMC/BLM procedure may be considered as the argument in favor of  validity of the two-fold representation for the NS Adler function proposed in  \cite{Cvetic:2016rot} at the four-loop level at least. 
 
In particular case $n_f=4$ the PT expression for $D_{NS}$-function has the following numerical form:
\begin{gather}
\label{DNS-n=5}
\hspace{-0.5cm}
D^{(M=5)}_{NS}(\overline{a}_s)\qquad =\qquad 1\qquad +\qquad \overline{a}_s\qquad +\qquad{\underbrace{1.5246}_{\sim \#\beta_0=1.4413}}\overline{a}^{\;^2}_s\qquad +
{\underbrace{2.7590}_{\sim \#\beta^2_0=13.4701 \atop \sim\#\beta^2_0+\#\beta_1+\#\beta_0=25.9817}}\hspace{-0.5cm}\overline{a}^{\;3}_s \\ \nonumber
\qquad + {\underbrace{27.3879}_{
 \sim \#\beta^3_0=19.7121 \atop \sim\#\beta^3_0+\#\beta_0\beta_1+\#\beta^2_0+\#\beta_2+\#\beta_1+\#\beta_0=-54.0900}}\hspace{-1.5cm}\overline{a}^{\;4}_s 
\qquad\qquad + {\underbrace{d_5}_{
 \sim \#\beta^4_0=579.0767 \atop \sim\#\beta^4_0+\#\beta^2_0\beta_1+\#\beta_0\beta_2+\#\beta^2_1+\#\beta_0\beta_1+\#\beta_3+\#\beta_2+\#\beta_1=746.8592}}\hspace{-2.45cm}\overline{a}^{\;5}_s \qquad + \qquad \mathcal{O}(\overline{a}^{\;6}_s),
\end{gather}
where numbers above the curly brackets equal to the total $\overline{a}^{\;M}_s$-contributions in the $M$-th order of PT and those below the brackets correspond either to the large-$\beta_0$ approximation or to the sum of the specified $\{\beta\}$-dependent terms shown in Eq.(\ref{DNS-beta-concrete}). It is seen from Eq.(\ref{DNS-n=5}) that in a number of instances the conformally-invariant terms $d_M[0]$ dominates the overall contribution of the $\{\beta\}$-expanded terms, changing even the total sign in numerical expression for $d_M$-coefficient. Since the generalizations of the BLM procedure allows to eliminate all $\{\beta\}$-dependent terms (\ref{DNS-beta-concrete}) by redefining the scale parameter in every order of PT
\cite{Grunberg:1991ac, Mikhailov:2004iq, Brodsky:2011ta, Brodsky:2013vpa, 
 Kataev:2014jba}, leaving only the conformally-invariant pieces $d_M[0]$ in expression for $d_M$, then it would be interesting to find out the possible asymptotic behavior of these conformal contributions at large $M$.
It is clear that in this case the leading large-$\beta_0$ or the subleading renormalon approximation has no impact on the asymptotics of the terms $d_M[0]$.  For the theory of renormalon studies of the perturbative series of various Euclidean quantities in the QCD, see e.g. works \cite{Zakharov:1992bx, Beneke:1994qe, Beneke:1998ui} and references therein.

In its turn, a non-renormalon mechanism analogous to the Lipatov technique \cite{Lipatov:1976ny}  has been studied previously in a number of works (see e.g. \cite{Itzykson:1977mf, Bogomolny:1978ft} and reviews \cite{Zinn-Justin:1980oco, Kazakov:1980rd}) to 
investigate the large order behavior of the PT series in the different quantum field models such as the conformal quenched QED. This approach, based on the expansion of the functional integral representation for the different Green functions at a non-trivial saddle points, also indicates the factorial growth of the higher order coefficients of the related PT series. The results of these works may be useful for careful studies of the possible asymptotics of the terms $d_M[0]$ at least in the case of the perturbative quenched QED as it was proposed in \cite{Itzykson:1977mf}.

\subsection{The case of the $e^+e^-$ annihilation R-ratio}

Physically the more important quantity is not the Adler $D$-function but the directly measurable Minkowskian
characteristics of the $e^+e^-$ annihilation process, namely $R$-ratio. It is related to the $D$-function
by the K\"allen-Lehmann integral representation (\ref{RtoDint-rel}). Taking into account the running effect of the 4-th order coupling constant $\overline{a}_s(s)$ in the Minkowskian region, which
may be obtained from the solution of Eq.(\ref{beta-exp}) by the transition $Q^2\rightarrow s$ from the Euclidean domain, one can arrive to the following  analytic correspondence:
\begin{eqnarray}
\label{correspondence}
R^{(M=5)}(\overline{a}_s)&=&D^{(M=5)}(\overline{a}_s)-\frac{\pi^2}{3}d_1\beta^2_0\overline{a}^{\;3}_s-\pi^2\bigg(d_2\beta^2_0+\frac{5}{6}d_1\beta_1\beta_0\bigg)\overline{a}^{\;4}_s \\ \nonumber
&-&\bigg[\pi^2\bigg(2d_3\beta^2_0+\frac{7}{3}d_2\beta_0\beta_1+\frac{1}{2}d_1\beta^2_1+d_1\beta_0\beta_2\bigg)-\frac{\pi^4}{5}d_1\beta^4_0\bigg]\overline{a}^{\;5}_s+\mathcal{O}(\overline{a}^{\;6}_s),
\end{eqnarray}
where additional terms proportional to the powers of $\pi^2$ are the analytic continuation effects from the Euclidean to Minkowski region. These $\pi^2$-effects may be found in the original works \cite{Kataev:1995vh, Bakulev:2010gm} up to the six-loop level. Thus, it follows from Eq.(\ref{correspondence}) that the difference between perturbative expressions for $R$-ratio and  Adler function starts to manifest itself from the 3-rd order of PT. 

Bearing in mind Eq.(\ref{correspondence}), one can conclude that the coefficients $r_n$ of the perturbative expression for the NS contribution to $R$-ratio in the $M$-th order of approximation
\begin{equation}
R^{(M)}_{NS}(\overline{a}_s)=1+\sum\limits_{n=1}^{M}r_n\overline{a}^{\;n}_s
\end{equation}
will have the similar $\{\beta\}$-expanded form as coefficients of the NS Adler function (\ref{d1}-\ref{d5}). However, some of them will get extra contributions, proportional to the analytical continuation $\pi^2$-effects:
\begin{eqnarray}
\label{fromDtoR}
&& r_3[2] = d_3[2] - \frac{\pi^2}{3} d_1[0], \qquad
r_4[2] = d_4[2] - \pi^2 d_2[0], \qquad r_4[1,1] = d_4[1,1] - \frac{5\pi^2}{6} d_1[0],  \\ \nonumber
&& r_4[3] = d_4[3] - \pi^2 d_2[1], \qquad
 r_5[2] = d_5[2] - 2 \pi^2 d_3[0], \qquad
 r_5[1,1] = d_5[1,1] - \frac{7\pi^2}{3} d_2[0],  \\
\nonumber
&& r_5[0,2] = d_5[0,2] -  \frac{\pi^2}{2} d_1[0], ~~~
 r_5[1,0,1] = d_5[1,0,1] - \pi^2 d_1[0], ~~~ r_5[3] = d_5[3] - 2 \pi^2 d_3[1], \\ \nonumber 
&& r_5[2,1] = d_5[2,1] - 2 \pi^2 d_3[0,1] - \frac{7\pi^2}{3} d_2[1], \qquad
r_5[4] = d_5[4] - 2 \pi^2 d_3[2] + \frac{\pi^4}{5} d_1[0].
\end{eqnarray}

All other $\{\beta\}$-expanded terms for $R_{NS}(\overline{a}_s)$, which do not receive the analytical continuation contributions at the $\mathcal{O}(\overline{a}^{\;5}_s)$ level, coincide with their $D_{NS}$-function counterparts, e.g. $r_3[1] = d_3[1]$, $r_5[0,0,1] = d_5[0,0,1]$, etc.

In the numerical form for the case of $SU(3)$ color gauge group these $\{\beta\}$-expanded terms read:
\begin{eqnarray}
\label{RNS-beta-concrete}
\nonumber
R^{(M=5)}_{NS}(\overline{a}_s)&=&1+\overline{a}_s+\bigg(0.6918\beta_0+\uuline{0.0833}\bigg)\overline{a}^{\;2}_s +\bigg(-0.1864\beta^2_0
+0.6918\beta_1+4.9402\beta_0-\uuline{23.2227}\bigg)\overline{a}^{\;3}_s \\ \nonumber
&+&\bigg(-4.6475\beta^3_0-2.0178\beta_0\beta_1
+16.8766\beta^2_0+
0.6918\beta_2+4.9402\beta_1 
-101.928\beta_0 \\ \nonumber
&+&\uuline{81.1571}+0.0802n_f\bigg)\overline{a}^{\;4}_s 
+\bigg(-11.0380\beta^4_0-23.0458\beta^2_0\beta_1
-3.6627\beta_0\beta_2-1.8314\beta^2_1 \\ \nonumber
&+& 33.4790\beta_0\beta_1 
+0.6918\beta_3+4.9402\beta_2-101.928\beta_1 \\
&+&\uwave{r_5[3]}\beta^3_0+\uwave{r_5[2]}\beta^2_0+\uwave{r_5[1]}\beta_0+\uwave{r_5[0]}\bigg)\overline{a}^{\;5}_s+\mathcal{O}(\overline{a}^{\;6}_s).
\end{eqnarray}

Note that the conformally-invariant terms $r_M[0]$ coincide with their analogs $d_M[0]$.

Mention also that starting with 3-loop level (see Eq.(\ref{RNS-beta-concrete})) the $\pi^2$-effects  contributing to $r_M[M-1]$ terms, which are responsible for the asymptotics in the large-$\beta_0$ approximation, lead to their negative values in comparison with the positive ones $d_M[M-1]$ in Eq.(\ref{DNS-beta-concrete}). Let us have a look what impact  will have these effects on the behavior of the PT series for $R$-ratio in a particular case $n_f=4$:
\begin{gather}
\label{RNS-n=5}
\hspace{-0.5cm}
R^{(M=5)}_{NS}(\overline{a}_s)\qquad =\qquad 1\qquad +\qquad \overline{a}_s\qquad + \qquad {\underbrace{1.5246}_{\sim \#\beta_0=1.4413}}\overline{a}^{\;^2}_s\qquad -
{\underbrace{11.5201}_{
 \sim \#\beta^2_0=-0.8090 \atop \sim\#\beta^2_0+\#\beta_1+\#\beta_0=11.7026}}\hspace{-0.5cm}\overline{a}^{\;3}_s \\ \nonumber
\qquad - {\underbrace{92.8916}_{
 \sim \#\beta^3_0=-42.0238 \atop \sim\#\beta^3_0+\#\beta_0\beta_1+\#\beta^2_0+\#\beta_2+\#\beta_1+\#\beta_0=-174.3695}}\hspace{-1.5cm}\overline{a}^{\;4}_s 
\qquad + {\underbrace{r_5}_{
 \sim \#\beta^4_0=-207.9340 \atop \sim\#\beta^4_0+\#\beta^2_0\beta_1+\#\beta_0\beta_2+\#\beta^2_1+\#\beta_0\beta_1+\#\beta_3+\#\beta_2+\#\beta_1=-646.3122}}\hspace{-2.45cm}\overline{a}^{\;5}_s \qquad + \qquad \mathcal{O}(\overline{a}^{\;6}_s).
\end{gather}

First of all, the values of $r_3$ and $r_4$ are negative and exceeds (in modulus) the positive values of $d_3$ and $d_4$ by almost 4 and 3.5 times correspondingly. Secondly, a value of the $\mathcal{O}(\overline{a}^{\;3}_s)$ large-$\beta_0$ contribution to $r_3$ is less nearly 14 times than $r_3$ itself, whereas $\beta^3_0$-term in $r_4$ makes up 45$\%$ of it. In the 3-rd and 4-th orders of PT the sum of $\{\beta\}$-dependent terms also differs substantially from the total corrections $r_3$ and $r_4$. This means that the conformal invariant terms $r_3[0]$ and $r_4[0]$ (\ref{RNS-n=5}) make a significant contributions to the overall corrections. Thirdly, in absolute value the large-$\beta_0$ contribution to $r_5$ is 3 times less than the sum of its known $\{\beta\}$-expanded terms. Both of them are comparable in order of magnitude to similar ones for the Adler function but are opposite in sign.

\subsection{The case of the Bjorken polarized sum rule}
\label{Bjsub}

Let us move on to the investigation of the second renorm-invariant quantity included in the generalized Crewther relation, namely to the Bjorken sum rule for the deep inelastic scattering of the polarized leptons on nucleons (\ref{BJPSR}) and more specifically to the NS Bjorken coefficient function $C_{NS}(\overline{a}_s)$. The dependence of the Bjorken coefficient function on the squared Euclidean momentum was extracted from the experimental data of  e.g. the CLAS Collaboration \cite{Deur:2014vea}  (Jefferson Lab) (see also the detailed experimentally-oriented review \cite{Deur:2018roz}), the COMPASS Collaboration \cite{COMPASS:2016jwv} (CERN)  and was utilized for  comparison with theoretical predictions of the QCD \cite{Kotlorz:2018bxp}. For the recent low energy investigations of the Bjorken sum rule, see \cite{Deur:2021klh}. 

The two-fold analog of Eq.(\ref{D-two-fold}) for the NS Bjorken function enables also to predict some its five-loop $\{\beta\}$-expanded terms and helps to carry out a partial verification of the fulfillment of the generalization of the Crewther relation in two-fold representation \cite{Kataev:2010dm, Kataev:2010du}.

In a full analogy with expressions (\ref{d1}-\ref{d5}), for coefficients $c_M$ of the NS Bjorken  function
\begin{equation}
\label{simC}
C^{(M=5)}_{NS}(\overline{a}_s)=1+c_1\overline{a}_s+c_2\overline{a}^{\;2}_s+c_3\overline{a}^{\;3}_s+c_4\overline{a}^{\;4}_s+c_5\overline{a}^{\;5}_s + \mathcal{O}(\overline{a}^{\;6}_s)
\end{equation}
one can write down:
\begin{subequations}
\begin{gather}
\label{c1}
c_1 = c_1[0], \\ 
\label{c2}
c_2 = \beta_0 c_2[1] + c_2[0],  \\ 
\label{c3}
c_3 = \beta_0^2 c_3[2] + \beta_1 c_3[0,1] + \beta_0 c_3[1] + c_3[0],  \\ 
\label{c4}
c_4 = \beta_0^3 c_4[3] + \beta_1 \beta_0 c_4[1,1] + \beta_2 c_4[0,0,1] + \beta_0^2 c_4[2] + \beta_1  c_4[0,1] + \beta_0 c_4[1] + c_4[0], 
\end{gather}
\begin{gather}
\label{c5}
c_5 = \beta^4_0 c_5[4] + \beta_1\beta^2_0 c_5[2,1] + \beta^3_0 c_5[3] + \beta_2 \beta_0 c_5[1,0,1] + \beta^2_1 c_5[0,2]  
+ \beta_1 \beta_0 c_5[1,1]  \\ \nonumber
+ \beta^2_0 c_5[2] + 
 \beta_3 c_5[0,0,0,1] + \beta_2 c_5[0,0,1] + \beta_1 c_5[0,1] + \beta_0 c_5[1] + c_5[0].
\end{gather}
\end{subequations}

Utilizing the analog of the representation (\ref{explicit}) for the non-singlet contribution to the Bjorken polarized sum rule, calculated analytically at the two-, three- and four-loop level in \cite{Gorishnii:1985xm, Larin:1991tj, Baikov:2010je} correspondingly, the authors of work \cite{Cvetic:2016rot} obtained all $\{\beta\}$-expanded terms in $c_2$, $c_3$ and $c_4$ coefficients in the $\MSbar$-scheme\footnote{Note that there is a misprint
in Eq.(14) of Ref.\cite{Cvetic:2016rot}. Instead of the 
$C_F$-coefficient $-151/24$ in the expression for $C_2(\overline{a}_s)$ should be $-115/24$ (see $c_3[2]$ in Table \ref{T-c1-3}).}. These results are given in Tables \ref{T-c1-3} and \ref{T-c4}.

\begin{table}[h!]
\renewcommand{\tabcolsep}{0.6cm} 
\renewcommand{\arraystretch}{1.7}
\centering
\begin{tabular}{|c|c|c|}
\hline
Coefficients              & Group structures &                                                      Numbers \\ \hline
$c_1[0]$                  & $C_F$            & $-\frac{3}{4}$                                         \\ \hline

 & $C^2_F$ & $\frac{21}{32}$                                       \\ \cline{2-3} 
 
\multirow{-2}{*}{$c_2[0]$}& $C_FC_A$         & $-\frac{1}{16}$                                        \\ \hline
$c_2[1]$                  & $C_F$            & $-\frac{3}{2}$                               \\ \hline
 & $C^3_F$          & $-\frac{3}{128}$                                     \\ \cline{2-3} 
 & $C^2_FC_A$       & $\frac{125}{256}-\frac{33}{16}\zeta_3$               \\ \cline{2-3} 

\multirow{-3}{*}{$c_3[0]$}  & $C_FC^2_A$       & $\frac{53}{192}+\frac{33}{16}\zeta_3$                \\ \hline
                                         
\multirow{2}{*}{$c_3[1]$} & $C^2_F$          & $\frac{349}{192}+\frac{5}{4}\zeta_3$                 \\ \cline{2-3} 
                          & $C_FC_A$         & $-\frac{155}{96}-\frac{9}{4}\zeta_3+\frac{5}{2}\zeta_5$ \\ \hline
$c_3[0,1]$                & $C_F$            & $-\frac{3}{2}$                               \\ \hline
$c_3[2]$                  & $C_F$            & $-\frac{115}{24}$                             \\ \hline
\end{tabular}
\captionsetup{justification=centering}
\caption{\label{T-c1-3} All terms included in the $\{\beta\}$-expansion of coefficients $c_1$, $c_2$ and $c_3$.}
\end{table}

\begin{table}[h!]
\renewcommand{\tabcolsep}{0.6cm} 
\renewcommand{\arraystretch}{1.5}
\centering
\begin{tabular}{|c|c|c|}
\hline
Coefficients                                                              & Group structures                      &                                                                                                                            Numbers \\ \hline
 & $C^4_F$                               & $-\frac{4823}{2048}-\frac{3}{8}\zeta_3$                                                                                      \\ \cline{2-3} 
 & $C^3_FC_A$                            & $\frac{605}{384}+\frac{469}{128}\zeta_3+\frac{165}{32}\zeta_5$                                                            \\ \cline{2-3} 
 & $C^2_FC^2_A$                          & $-\frac{11071}{4608}-\frac{695}{128}\zeta_3-\frac{165}{64}\zeta_5$                                                            \\ \cline{2-3} 
 & $C_FC^3_A$                            & $\frac{30863}{36864}+\frac{147}{128}\zeta_3-\frac{165}{64}\zeta_5$                                                         \\ \cline{2-3} 
 & $\frac{d^{abcd}_Fd^{abcd}_A}{d_R}$    & $-\frac{3}{16}+\frac{1}{4}\zeta_3+\frac{5}{4}\zeta_5$                                                                        \\ \cline{2-3} 
\multirow{-6}{*}{$c_4[0]$} & $\frac{d^{abcd}_Fd^{abcd}_F}{d_R}n_f$ & $\frac{13}{16}+\zeta_3-\frac{5}{2}\zeta_5$                                                                                 \\ \hline
                                                                          & $C^3_F$                               & $-\frac{997}{384}-\frac{481}{32}\zeta_3+ \frac{145}{8}\zeta_5$ \\ \cline{2-3} 
                                                                          & $C^2_FC_A$                            & $\frac{85801}{4608}+\frac{169}{24}\zeta_3-\frac{365}{48}\zeta_5-\frac{105}{4}\zeta_7$                 \\ \cline{2-3} 
\multirow{-3}{*}{$c_4[1]$}                                                & $C_FC^2_A$                            & $-\frac{931}{768}+\frac{955}{192}\zeta_3+\frac{895}{96}\zeta_5+\frac{105}{16}\zeta_7$                \\ \hline
                                                                          & $C^2_F$                               & $\frac{349}{192}+\frac{5}{4}\zeta_3$                                                                                       \\ \cline{2-3} 
\multirow{-2}{*}{$c_4[0,1]$}                                              & $C_FC_A$                              & $-\frac{155}{96}-\frac{9}{4}\zeta_3+\frac{5}{2}\zeta_5$                                                                       \\ \hline
                                                                          & $C^2_F$                               & $\frac{261}{64}+\frac{87}{8}\zeta_3$                                                 \\ \cline{2-3} 
\multirow{-2}{*}{$c_4[2]$}                                                & $C_FC_A$                              & $-\frac{3151}{256}-\frac{43}{16}\zeta_3-\frac{3}{2}\zeta^2_3+\frac{15}{4}\zeta_5$                                          \\ \hline
$c_4[0,0,1]$                                                              & $C_F$                                 & $-\frac{3}{2}$                                                                                                     \\ \hline
$c_4[1,1]$                                                                & $C_F$                                 & $-\frac{115}{12}$                                                                                                   \\ \hline
$c_4[3]$                                                                  & $C_F$                                 & $-\frac{605}{36}$                                                                            \\ \hline
\end{tabular}
\captionsetup{justification=centering}
\caption{\label{T-c4} All terms included in the $\{\beta\}$-expansion of coefficient $c_4$.}
\end{table}

As in the case of the $d_4[0]$-term for the NS Adler function, the conformal invariant coefficients $c_4[0]$  also contain $d^{abcd}_Fd^{abcd}_A/d_R$ and $d^{abcd}_Fd^{abcd}_Fn_f/d_R$ contributions of the light-by-light scattering type, exactly the same as in $d_4[0]$ but with the opposite sign. Thus, they are safely canceled out due to the conformal symmetry relations in the product of the NS Adler and Bjorken functions in the generalized Crewther relation as was observed previously in \cite{Kataev:2010du}.

Acting in full analogy with the case of the Adler function in Subsection \ref{SubSecAdler}, we obtain the similar 5-th order $\{\beta\}$-expanded analogs of the Bjorken polarized sum rule, presented in Table \ref{T-c5}.

\begin{table}[h!]
\renewcommand{\tabcolsep}{0.6cm} 
\renewcommand{\arraystretch}{1.8}
\centering
\begin{tabular}{|c|c|c|}
\hline 
~~~Coefficients~~~                 & ~~~Group structures~~~ &                                                                                                            Numbers \\ \hline
\multirow{3}{*}{$c_5[0,1]$}   & 
$C^3_F$          &  $-\frac{997}{384}-\frac{481}{32}\zeta_3+\frac{145}{8}\zeta_5$  \\  \cline{2-3}
  & $C^2_FC_A$       & ~~$\frac{85801}{4608}+\frac{169}{24}\zeta_3-\frac{365}{48}\zeta_5-\frac{105}{4}\zeta_7$~~ \\  \cline{2-3} 
                              & $C_FC^2_A$       & $-\frac{931}{768}+\frac{955}{192}\zeta_3+\frac{895}{96}\zeta_5+\frac{105}{16}\zeta_7$       \\   \hline
\multirow{2}{*}{$c_5[0,0,1]$} & $C^2_F$          & $\frac{349}{192}+\frac{5}{4}\zeta_3$                                                                       \\  \cline{2-3} 
                              & $C_FC_A$         & $-\frac{155}{96}-\frac{9}{4}\zeta_3+\frac{5}{2}\zeta_5$                                                       \\   \hline
$c_5[0,0,0,1]$                & $C_F$            & $-\frac{3}{2}$
\\ \hline                                                        
\multirow{2}{*}{$c_5[1,1]$}   & $C^2_F$          & $\frac{261}{32}+\frac{87}{4}\zeta_3$                                            \\    \cline{2-3} 
                              & $C_FC_A$         & $-\frac{3151}{128}-\frac{43}{8}\zeta_3-3\zeta^2_3+\frac{15}{2}\zeta_5$                                     \\    \hline
$c_5[0,2]$                    & $C_F$            & $-\frac{115}{24}$                                                                                   \\     \hline
$c_5[1,0,1]$                  & $C_F$            & $-\frac{115}{12}$                                                                                   \\      \hline
$c_5[2,1]$                    & $C_F$            & $-\frac{605}{12}$                                                           \\      \hline
$c_5[4]$                    & $C_F$            & $-\frac{1867}{24}$                                                          
\\     \hline
\end{tabular}
\captionsetup{justification=centering}
\caption{\label{T-c5} All extracted terms included in the $\{\beta\}$-expansion of coefficient $c_5$.}
\end{table}

Note that the leading renormalon contribution $c_5[4]$ was extracted by us from the results of work \cite{Broadhurst:1993ru}. Thus, in the case of the NS Bjorken function with help of its two-fold representation at the five-loop level we can also define 8 out of 12 coefficients in the $\{\beta\}$-expansion of $c_5$  (\ref{c5}). 

In the numerical form for the case of $SU(3)$ group the presented $\{\beta\}$-expanded coefficients read:
\begin{eqnarray}
\label{CNS-beta-concrete}
\nonumber
C^{(M=5)}_{NS}(\overline{a}_s)&=&1-\overline{a}_s+\bigg(-2\beta_0+\uuline{0.9167}\bigg)\overline{a}^{\;2}_s +\bigg(-6.3889\beta^2_0
-2\beta_1-1.0048\beta_0+\uuline{22.3894}\bigg)\overline{a}^{\;3}_s \\ \nonumber
&+&\bigg(-22.4074\beta^3_0-12.7778\beta_0\beta_1
-24.7824\beta^2_0-2\beta_2-1.0048\beta_1 
+209.4096\beta_0 \\ \nonumber
&-&\uuline{126.8456}-0.0802n_f\bigg)\overline{a}^{\;4}_s 
+\bigg(-103.7222\beta^4_0-67.2222\beta^2_0\beta_1
-12.7778\beta_0\beta_2 \\ \nonumber
&-& 6.3889\beta^2_1-49.5649\beta_0\beta_1 -2\beta_3-1.0048\beta_2
+209.4096\beta_1 \\
&+&\uwave{c_5[3]}\beta^3_0+\uwave{c_5[2]}\beta^2_0+\uwave{c_5[1]}\beta_0+\uwave{c_5[0]}\bigg)\overline{a}^{\;5}_s+\mathcal{O}(\overline{a}^{\;6}_s).
\end{eqnarray}

Here 4 wavy terms out of 12 possible ones remain undetermined (but with the fixed $\zeta_4$-contributions to $c_5[0], c_5[1], c_5[2]$ -- see extra clarifications below). One should mention that the values of the conformal-invariant double underlined terms $c_2[0]=0.9167$ and $c_3[0]=22.3894$ (\ref{CNS-beta-concrete}) are in  agreement with those, obtained in \cite{Kataev:1992jm} for the Gross-Llewellyn-Smith sum rule of the neutrino-nucleon deep-inelastic scattering,  whose  non-singlet part is identical to the NS Bjorken polarized sum rule. In this quoted work all $n_f$-dependent terms are absorbed into the $\overline{a}_s$-dependent scale in each considered order up to the three-loop level.

For the specific case $n_f=4$ the PT expression for the NS Bjorken polarized sum rule has the following numerical form:
\begin{gather}
\label{CNS-n=5}
\hspace{-0.5cm}
C_{NS}(\overline{a}_s)\qquad =\qquad 1\qquad -\qquad \overline{a}_s\qquad -\qquad{\underbrace{3.2500}_{\sim \#\beta_0=-4.1667}}\overline{a}^{\;^2}_s\qquad -
{\underbrace{13.8502}_{
 \sim \#\beta^2_0=-27.7296 \atop \sim\#\beta^2_0+\#\beta_1+\#\beta_0=-36.2396}}\hspace{-0.5cm}\overline{a}^{\;3}_s \\ \nonumber
\qquad - {\underbrace{102.4015}_{
 \sim \#\beta^3_0=-202.6132 \atop \sim\#\beta^3_0+\#\beta_0\beta_1+\#\beta^2_0+\#\beta_2+\#\beta_1+\#\beta_0=24.7649}}\hspace{-1.5cm}\overline{a}^{\;4}_s 
\qquad\qquad + {\underbrace{c_5}_{
 \sim \#\beta^4_0=-1953.9200 \atop \sim\#\beta^4_0+\#\beta^2_0\beta_1+\#\beta_0\beta_2+\#\beta^2_1+\#\beta_0\beta_1+\#\beta_3+\#\beta_2+\#\beta_1=-2853.3681}}\hspace{-2.45cm}\overline{a}^{\;5}_s \qquad + \qquad \mathcal{O}(\overline{a}^{\;6}_s),
\end{gather}

As seen from Eq.(\ref{CNS-n=5}) the contributions, predicted by the large-$\beta_0$ approximation, yield plausible estimates for real values of the coefficients $c_M$. This observation will also hold when $n_f=5$. Unlike the Bjorken polarized sum rule, for the case of the NS Adler function the large-$\beta_0$ approximation works worse starting from the three-loop level at $n_f=5$. This fact is in agreement with the renormalon calculus \cite{Beneke:1998ui}.

Finishing this section, we conclude that the joint usage of the two-fold representation for the NS Adler function, $R$-ratio and Bjorken polarized sum rule with their $\{\beta\}$-expansion form at the five-loop order allows to fix their previously unknown definite $\beta$-dependent non-conformal invariant terms of the 5-th order of PT. Knowledge of the explicit form of these contributions may be important say for the implementation of more refined estimates or calculations of these corrections. Moreover, in the 5-th order of PT there is an additional approach enabling to fix the Riemann $\zeta_4$-contributions and their Lie group structures
in 3 out of 4 remaining undetermined $\{\beta\}$-expanded terms (see \cite{Goriachuk:2020oah}). Let us consider this statement in more detail.

\section{Riemann $\zeta_4$-contributions to $d_5$ and $c_5$-coefficients}

The remarkable all-order no-$\pi$ theorem was proved in the work \cite{Baikov:2018wgs} and its consequences were summarized in Ref.\cite{Baikov:2019zmy} for the generic one-charge theory and QCD in particular. For instance, it guarantees the $\pi$-independence of the QCD {\it{scale-invariant}} five-loop function corresponding to an every renormalized Green function or a two-point correlator defined in the Euclidean space. In context of the $e^+e^-$ annihilation process this fact  provides a link \cite{Baikov:2018wgs} between the $\pi^2$-dependent $\mathcal{O}(\overline{a}^{\;5}_s)$-contribution to the NS Adler function and  the five-loop coefficient $\beta_4$. As known from explicit calculations \cite{Baikov:2016tgj, Herzog:2017ohr, Luthe:2017ttc}, the coefficient $\beta_4$ 
contains only one even Riemann zeta-function, namely $\zeta_4=\pi^4/90$ term. Thus, the $\MSbar$-scheme coefficient $d_5$ will also contain this transcendental constant of weight 4. It completely explains the conjecture  \cite{Jamin:2017mul, Davies:2017hyl} on the first appearance of $\zeta_4$-terms in the PT expression for the NS Adler function in the fifth order and on 
the absence of even Riemann zeta-functions in all higher-order PT corrections to the RG Euclidean invariant  quantities, calculated in the renormalization effective $C$-scheme.

This specific QCD scheme was proposed in \cite{Boito:2016pwf}. Its characteristic feature is that the $\mu^2_C$-scale evolution of the coupling constant $a^C_s$ is governed by the $\beta^C$-function, depending on two coefficients $\beta_0$ and $\beta_1$ in the following way:
\begin{equation}
\label{beta-C}
\beta^C(a^C_s)=\mu^2_C\frac{\partial a^C_s}{\partial \mu^2_C}=\frac{-\beta_0(a^C_s)^2}{1-(\beta_1/\beta_0)a^C_s}.
\end{equation}

The transition from the $C$-scheme to the $\MSbar$-scheme can be implemented with help of the finite renormalization \cite{Boito:2016pwf, Jamin:2017mul}.
Along with the results of direct examinations \cite{Baikov:2018wgs}, this indirect
procedure also enables to unambiguously restore the explicit form of the coefficients and group structures proportional to $\zeta_4$-terms in the $\MSbar$-scheme expression for $D_{NS}$ and $C_{NS}$. In order to obtain their form for the case of the generic simple gauge group, we will adhere to the more transparent and evident in our opinion second way, supplemented by the effective charges approach \cite{Grunberg:1982fw, Kataev:1995vh}. Naturally, this roundabout way confirms the results following from the work \cite{Baikov:2018wgs}.

Using the effective charges approach, one can get the following five-loop coefficient of the effective $\beta$-function, constructed for the effective coupling of the Adler function:
\begin{eqnarray}
\label{beta4eff}
\beta^{eff}_4&=&\beta_4-3\beta_3\frac{d_2}{d_1}+\beta_2\bigg(4\frac{d^2_2}{d^2_1}-\frac{d_3}{d_1}\bigg)+\beta_1\bigg(\frac{d_4}{d_1}-2\frac{d_2d_3}{d^2_1}\bigg) \\ \nonumber
&+&\beta_0\bigg(3\frac{d_5}{d_1}-12\frac{d_2d_4}{d^2_1}-5\frac{d^2_3}{d^2_1}+28\frac{d^2_2d_3}{d^3_1}-14\frac{d^4_2}{d^4_1}\bigg).
\end{eqnarray}

To obtain the analog of this expression for the Bjorken polarized sum rule it is necessary to replace $d_n\rightarrow c_n$.

Formula (\ref{beta4eff}) is valid both the $C$-scheme and $\MSbar$-one. On the one hand, in the $C$-scheme the coefficients $d_n$ do not contain $\zeta_4$-contributions \cite{Jamin:2017mul, Baikov:2018wgs, Baikov:2019zmy} and $\beta_n$ also do not include them because they are expressed trough $\beta_0$ and $\beta_1$ only (\ref{beta-C}). Thus, for the case of the $C$-scheme $\beta^{eff}_4$-coefficient is free of $\zeta_4$. On the other hand, in the $\MSbar$-scheme the r.h.s. of Eq.(\ref{beta4eff}) contains this Riemann function in the five-loop coefficient $\beta_4$ and coefficient $d_5$ only. The crucial point here is the scheme invariance of the coefficients of the $\beta^{eff}$-function \cite{Stevenson:1981vj}, constructed by means of the effective charges approach within the gauge-independent renormalization schemes. Therefore, the expressions for $\beta^{eff}_4$, obtained from the $C$- and $\MSbar$-schemes, should coincide. This leads to the following equality, relating $\zeta_4$-contributions to the $\MSbar$ coefficients $\beta_4$ and $d_5$, which was also derived previously in \cite{Baikov:2018wgs}:
\begin{equation}
\label{relation-zeta4}
0=\beta^{(\zeta_4)}_4+3\beta_0\frac{d^{(\zeta_4)}_5}{d_1} \qquad \Rightarrow \qquad  d^{(\zeta_4)}_5=-d_1\frac{\beta^{(\zeta_4)}_4}{3\beta_0}.
\end{equation}

Moreover, it follows from \cite{Baikov:2018wgs} that the $\zeta_4$-part of the 5-loop $\MSbar$-scheme coefficient $\beta_4$ is divided by $\beta_0$ without reminder and is expressed through $\zeta_3$-part of $\beta_3$: 
\begin{equation}
\label{ratio-4-0}
\frac{\beta^{(\zeta_4)}_4}{\zeta_4\beta_0}=-\frac{9}{8}\frac{\beta^{(\zeta_3)}_3}{\zeta_3}.
\end{equation}

Substituting ratio (\ref{ratio-4-0}) into Eq.(\ref{relation-zeta4}), one can obtain  
\begin{equation}
\label{d5-zeta-3}
d^{(\zeta_4)}_5=\frac{3}{8}d_1\frac{\zeta_4}{\zeta_3}\beta^{(\zeta_3)}_3.
\end{equation}

Taking into account Eq.(\ref{d5-zeta-3}), the explicit expression for the $\zeta_3$-part of $\beta_3$ 
 \cite{vanRitbergen:1997va, Czakon:2004bu}
in the generic simple gauge group and replacing there the factor $N_A$, containing in combinations with the group structures $d^{abcd}_Fd^{abcd}_F$, $d^{abcd}_Fd^{abcd}_A$, $d^{abcd}_Ad^{abcd}_A$ and being equal to the dimension of the adjoint representation ($N_A=N^2_c-1$ for the $SU(N_c)$-group), with help of the equality $T_FN_A=C_Fd_R$, we find $\zeta_4$-contributions to the coefficients $d_5$ and $c_5$ in a more traditional form, involving $d_R$ rather than $N_A$ (see also \cite{Goriachuk:2020oah}): 
\begin{eqnarray}
\label{d5-c5-z4}
d^{(\zeta_4)}_5=-c^{(\zeta_4)}_5&=&\zeta_4\bigg(-\frac{11}{2048}C_FC^4_A+\frac{11}{256}C^3_FC_AT_Fn_f-\frac{41}{512}C^2_FC^2_AT_Fn_f \\ \nonumber
&+&\frac{51}{1024}C_FC^3_AT_Fn_f
-\frac{11}{128}C^3_FT^2_Fn^2_f+\frac{7}{128}C^2_FC_AT^2_Fn^2_f+\frac{7}{256}C_FC^2_AT^2_Fn^2_f \\ \nonumber
&+&\frac{33}{128}\frac{d^{abcd}_Ad^{abcd}_A}{d_R}T_F
-\frac{39}{64}\frac{d^{abcd}_Fd^{abcd}_A}{d_R}T_Fn_f+\frac{3}{16}\frac{d^{abcd}_Fd^{abcd}_F}{d_R}T_Fn^2_f\bigg).
\end{eqnarray}

In the case of $N_c=3$ this expression takes the following simple form
\begin{equation}
\label{d5-c5-z4-N=3}
d^{(\zeta_4)}_5=-c^{(\zeta_4)}_5=\zeta_4\bigg(\frac{2673}{512}-\frac{1627}{4608}n_f+\frac{809}{6912}n^2_f\bigg),
\end{equation}
which coincides with the result for $\zeta_4$-corrections to $D$-function of the $SU(3)$ QCD, obtained in \cite{Jamin:2017mul}.

Let us turn to the definition of the 5-th order $\{\beta\}$-expanded terms in the NS Adler function and Bjorken polarized sum rule, containing $\zeta_4$-contributions. As it is seen from Eq.(\ref{d5-c5-z4}) the coefficients $d^{(\zeta_4)}_5$ and $c^{(\zeta_4)}_5$ hold quadratic contributions in $n_f$. Relying on the $\{\beta\}$-expansion representation for $d_5$ (\ref{d5}) and $c_5$ (\ref{c5}), we could conclude that theoretically the decomposition of $d^{(\zeta_4)}_5$ and $c^{(\zeta_4)}_5$ may be in terms $\beta^2_1$, $\beta_1\beta_0$, $\beta^2_0$, $\beta_2$, $\beta_1$, $\beta_0$ and $\beta$-independent one. However, contributions proportional to $\beta^2_1$, $\beta_1\beta_0$, $\beta_2$ and $\beta_1$, namely $d_5[0,2]$, $d_5[1,1]$, $d_5[0,0,1]$ and $d_5[0,1]$ (and their Bjorken counterparts), we have already determined (see Table \ref{T-d5}, Eq.(\ref{DNS-beta-concrete}) and Table \ref{T-c5}, Eq.(\ref{CNS-beta-concrete})) with help of the two-fold representation. These terms are free of $\zeta_4$-function. Therefore, we may write down the following expansion:
\begin{equation}
\label{d5-z4-beta}
d_5^{(\zeta_4)} = \beta^2_0 d_5^{(\zeta_4)}[2] + \beta_0 d_5^{(\zeta_4)}[1] + d_5^{(\zeta_4)}[0]. 
\end{equation}

Following the idea, outlined in \cite{Goriachuk:2020oah}, i.e. decomposing terms $d_5^{(\zeta_4)}[2]$, $d_5^{(\zeta_4)}[1]$, $d_5^{(\zeta_4)}[0]$ into the
Lie group structures, using the result (\ref{d5-c5-z4}) and the explicit form $\beta_0=11C_A/12-T_Fn_f/3$, we arrive to the system of linear equations, whose exact solution is: 
\begin{subequations}
\begin{eqnarray}
\label{d5-z4-0}
d_5^{(\zeta_4)}[0] = - c_5^{(\zeta_4)}[0] &=& \zeta_4 \bigg(\frac{693}{2048}C_FC^4_A +
\frac{99}{512} C^2_FC^3_A - \frac{1089}{2048}C^3_FC^2_A   \\ \nonumber
&+& \frac{33}{128} \frac{d_A^{abcd}d_A^{abcd}}{d_R} T_F - \frac{429}{256}\frac{d_F^{abcd}d_A^{abcd}}{d_R} C_A +
\frac{33}{64} \frac{d_F^{abcd}d_F^{abcd}}{d_R}C_A n_f \bigg),   \\
\label{d5-z4-1}
d_5^{(\zeta_4)}[1] = - c_5^{(\zeta_4)}[1] &=& \zeta_4 \bigg(-\frac{615}{1024} C_FC^3_A -
\frac{339}{512} C^2_FC^2_A + \frac{165}{128} C^3_FC_A   \\ \nonumber
&+& \frac{117}{64}\frac{d_F^{abcd}d_A^{abcd}}{d_R} - \frac{9}{16}\frac{d_F^{abcd}d_F^{abcd}}{d_R} n_f\bigg),  \\
\label{d5-z4-2}
d_5^{(\zeta_4)}[2] = - c_5^{(\zeta_4)}[2] &=& \zeta_4 \bigg(\frac{63}{256} C_FC^2_A +
\frac{63}{128} C^2_FC_A - \frac{99}{128} C^3_F\bigg). 
\end{eqnarray}
\end{subequations}

These results reproduce those presented in  \cite{Goriachuk:2020oah}. However, in this quoted work the two-fold representation was not applied and the absence in the expansion (\ref{d5-z4-beta}) of other $\{\beta\}$-dependent terms, different from $\beta_0$ and $\beta^2_0$ ones, was shown there in more complicated way. 

The expressions (\ref{d5-c5-z4}) and (\ref{d5-z4-0}-\ref{d5-z4-2}) demonstrate what the Lie group structures will be exactly contained in the 5-th order corrections to the NS Adler function and NS Bjorken polarized sum rule and in their some $\{\beta\}$-expanded terms. Note that in contrast to the four-loop approximation, at the $\mathcal{O}(\overline{a}^{\;5}_s)$ level the light-by-light scattering effects reveal themselves not only in the conformally-invariant terms $d_5[0]$ and $c_5[0]$, but also in the
proportional to $\beta_0$ ones, namely in $d_5[1]$ and $c_5[1]$. This partially occurs due to the insertion of fermion loop into the external gluon line, entering to the light-by-light scattering subgraph.

Let us make a few remarks about the nature of reductions of certain $\zeta_4$-contributions in the generalized Crewther relation. 
Since $d_1=-c_1$, then accordingly to Eq.(\ref{d5-zeta-3}) and its Bjorken analogue, 
the sum $d^{(\zeta_4)}_5+c^{(\zeta_4)}_5=0$. Thus, the $\zeta_4$-terms are mutually canceled out in this sum. As known in the class of the gauge-invariant ${\rm{MS}}$-like renormalization schemes the coefficients of the RG $\beta$-function do not change upon the scale transformations $\mu'=\lambda\mu$, where $\lambda$ is the constant. The one-loop coefficients $d_1$ and $c_1$ are also scale-independent. Therefore,  in accordance with Eq.(\ref{d5-zeta-3}) the terms, proportional to $\zeta_4$ in $d_5$ and $c_5$, are scale-independent as well\footnote{It is interesting to trace explicitly how they will (not) change upon the transformations of the Poicar\'e group (rotations, Lorenz boosts, translations, reflections) and upon the special conformal transformations. The representation of the conformal transformations in the momentum space was considered e.g. in \cite{Maglio:2021yaq}.}
 and the expression (\ref{d5-c5-z4}) stays valid in  all class of the ${\rm{MS}}$-like schemes. Thus, these $\zeta_4$-terms are invariant under the scale transformations or, in other words, under dilatations, which are a particular case of the conformal transformations. 

In this regard, considering the relation
\begin{eqnarray}
\label{crew-d5-c5}
d_5+c_5+d_1c_4+c_1d_4+d_2c_3+c_2d_3=-\beta_3K_1-\beta_2K_2-\beta_1K_3-\beta_0K_4,
\end{eqnarray}
following from the 5-loop generalization of the Crewther relation in the one-fold representation (see  e.g. \cite{Garkusha:2018mua} or 5-loop analog of Eq.(\ref{BK})), one can conclude that the 4-th coefficient $K_4$, entering in the conformal symmetry breaking term $(\beta^{(N=4)}(\overline{a}_s)/\overline{a}_s)K^{(N=4)}(\overline{a}_s)$, does not contain $\zeta_4$-terms. Indeed, in the sum $d_5+c_5$ they are mutually canceled out and the rest coefficients $d_n$, $c_n$ with $n=1,2,3,4$, $K_1$, $K_2$, $K_3$ and $\beta_k$ with $k=0,1,2,3$ are free of $\zeta_4$. This fact is the consequence of the scale invariance.

In the conformal invariant limit when all $\beta_k=0$, the relation (\ref{crew-d5-c5}) takes the following form
\begin{eqnarray}
\label{crew-d5-c5-conf}
d_5[0]+c_5[0]+d_1[0]c_4[0]+c_1[0]d_4[0]+d_2[0]c_3[0]+c_2[0]d_3[0]=0,
\end{eqnarray}
which can be derived from more general expression of \cite{Kataev:2010du}.

Here $\zeta_4$-contributions are contained only in $d_5[0]$, $c_5[0]$  and are mutually canceled out in this sum (\ref{d5-z4-0}). Thus, the disappearance of these terms in the mentioned sum is in agreement with 
the conformal limit of the generalized Crewther identity. This means that the cancellation of $\zeta_4$-terms in $d^{(\zeta_4)}_5[0]+c^{(\zeta_4)}_5[0]$ is the consequence of the conformal symmetry which is more general compared to the scale symmetry.

\section{Discussion and outlook}
\label{Discussion}

Since there is currently no consensus on the order of separation of conformally invariant contributions in the perturbative expressions for the NS Adler function and Bjorken polarized sum rule (see e.g. Refs.\cite{Brodsky:1982gc, Grunberg:1991ac,  Mikhailov:2004iq, Brodsky:2011ta, Brodsky:2013vpa,  Kataev:2014jba, Kataev:2016aib, Cvetic:2016rot,  Mikhailov:2016feh, Wu:2019mky, Huang:2020skl} on this topic), we will dwell on this issue in more detail and try to make convincing arguments in favor of the correct in our opinion procedure for their extraction. 

As we have seen in the case of the NS Adler function in the $SU(3)$ QCD the two-, three- and four-loop conformally invariant terms are $d_2[0]=0.0833$, $d_3[0]=-23.2227$, $d_4[0]=81.1571+0.0802n_f$ (\ref{DNS-beta-concrete}). The small $n_f$-dependent contribution to $d_4[0]$ is a manifestation of the light-by-light scattering effects. These values are also presented in \cite{Cvetic:2016rot} and are
in agreement with results of application of the PMC/BLM procedure\footnote{Note that the additional ambiguities of the PMC method, not discussed in this work, were considered in Ref.\cite{Chawdhry:2019uuv}. }
in works \cite{Grunberg:1991ac}, \cite{Brodsky:2011ta} and with expression (A5) from Appendix A of Ref.\cite{Brodsky:2013vpa}. But they differ from those, given e.g. in the main part of \cite{Brodsky:2013vpa} and other related works \cite{Wu:2019mky, Huang:2020skl}. In these works the representation of the $D$-function in terms of the photon anomalous dimension $\gamma(a_s)$ and the hadronic vacuum polarization function $\Pi(L=\ln\mu^2/Q^2, a_s)$ is used, first studied in Refs.\cite{Chetyrkin:1980sa, Chetyrkin:1980pr, Baikov:2012zm}:
\begin{equation}
\label{D-anomalous}
D(L, a_s)=\gamma(a_s)-\beta(a_s)\frac{\partial }{\partial a_s}\Pi(L, a_s).
\end{equation}

This relation follows from the definition (\ref{RtoDint-rel}) and the inhomogeneous RG equation for $\Pi(L, a_s)$
\begin{equation}
\label{inhomog}
\mu^2\frac{d}{d\mu^2}\Pi(L, a_s)\bigg|_{L=0}=\bigg(\mu^2\frac{\partial}{\partial \mu^2}+\beta(a_s)\frac{\partial}{\partial a_s}\bigg)\Pi(L, a_s)\bigg|_{L=0}=\gamma(a_s).
\end{equation}

It was mentioned in \cite{Brodsky:2013vpa, Wu:2019mky} that the anomalous dimension $\gamma(a_s)$ is associated with the renormalization of the QED coupling only and is not related to the running of the strong coupling constant (for details see section IV of the first cited work and 4.3 of the second one). This RG function is treated as conformal contribution during the PMC scale setting analysis and is not decomposed in the $\{\beta\}$-expanded terms in these quoted works. We state that these considerations are not true. Indeed, $\gamma(a_s)$-function is \textit{inseparably related} with the renormalization of the QCD charge. This fact has already been discussed in \cite{Kataev:2014jba} and studied in
\cite{Kataev:2016aib}. Let us focus on this issue in more detail. 

As known the two-point photon correlator, renormalized by the QCD radiative corrections, is
\begin{equation}
\label{correlator}
G_{\mu\nu}(q)=\frac{i}{q^2}\bigg[\bigg(-g_{\mu\nu}+\frac{q_\mu q_\nu}{q^2}\bigg)\frac{1}{1+a\Pi(L, a_s)}-\xi \frac{q_\mu q_\nu}{q^2}\bigg],
\end{equation}
where $a=a(\mu^2)=\alpha(\mu^2)/\pi$ is the  electromagnetic coupling renormalized by the strong interactions only, $\xi$ is the gauge covariant parameter. 

Taking into account the renormalization prescription for the bare and renormalized photon fields, namely $A^{\mu}_B=\sqrt{Z_{ph}}A^{\mu}$, and the non-renormalizability of the longitudinal part of the propagator, one can obtain the following relation between the renormalized $\Pi$ and bare $\Pi_B$ polarization functions
\begin{equation}
\label{link}
1+a\Pi(L, a_s)=Z_{ph}\bigg(1+a_B\Pi_B(L, a_{sB})\bigg),
\end{equation}
which was presented in \cite{Chetyrkin:1980sa, Chetyrkin:1980pr}. Due to the gauge invariance of the $\MSbar$-scheme neither $\Pi$ nor $\Pi_B$ depend on $\xi$. The 3-rd and 4-th order corrections to $\Pi$ were presented in \cite{Baikov:2012zm}. The condition of independence of the bare coupling constant $a_{sB}$ on the scale $\mu$ leads to the following relation:
\begin{eqnarray}
\label{asB}
a_{s,B}&=&\mu^{2\varepsilon}a_s\;{{\rm{exp}}}\;\bigg(-\int\limits_0^{a_s} \frac{dx}{x}\frac{\beta(x)}{\beta(x)-\varepsilon x}\bigg) \\ \nonumber
&=&\mu^{2\varepsilon}\bigg(a_s-\frac{\beta_0}{\varepsilon}a^2_s+\left(\frac{\beta^2_0}{\varepsilon^2}
-\frac{\beta_1}{2\varepsilon}\right)a^3_s  
-\left(\frac{\beta^3_0}{\varepsilon^3}-\frac{7\beta_1\beta_0}{6\varepsilon^2}
+\frac{\beta_2}{3\varepsilon}\right)a^4_s +\mathcal{O}(a^5_s)\bigg).
\end{eqnarray}

Here in the definition of $\beta$-function we retain $\varepsilon$-contribution i.e. $\beta(a_s)=-\varepsilon a_s-\sum\limits_{i\geq 0}\beta_i a^{\;i+2}_s$, where $\varepsilon=(4-d)/2$ is the parameter of the dimensional regularization.

Owing to the Ward identity with the omitted prefactor $\mu^{2\varepsilon}$, we have $a=Z_{ph}a_B$. For the QCD coupling one can write $a_s=Z^{-1}_{a_s}a_{sB}$. In the class of the ${\rm{MS}}$-like schemes $Z_{ph}$ reads:
\begin{equation}
\label{epsilon}
Z_{ph}=1+a\cdot Z(a_s)=1+a\cdot\sum\limits_{p\geq 1}a^{p-1}_s\sum\limits_{k=1}^p \frac{Z_{p,-k}}{\varepsilon^k}.
\end{equation}

\begin{figure}[h!]
\centering
\scalebox{0.9}{
\includegraphics[width=\textwidth]{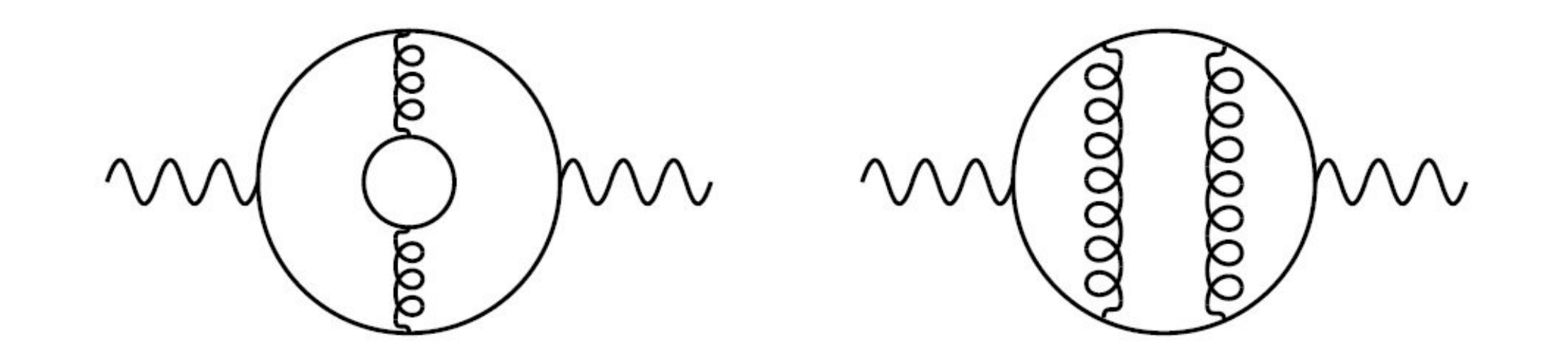}
}
\centering
\captionsetup{justification=centering}
\caption{The specific diagrams, contributing to the $\beta_0$-dependent (within the Naive-Nonabelianization procedure) and conformally invariant terms of $d_2$-coefficient correspondingly.}
\label{diagrams}
\end{figure}

Here the QED coupling $a$ is included in $Z_{ph}$ due to the Feynman diagram with fermion loop and two external photon legs. The remaining $a_s$-corrections arise from the transmission of gluon propagators with their internal inserts in the mentioned fermion loop. Naturally, part of these contributions is related to the renormalization of the QCD charge. For instance, see Figure \ref{diagrams}, where the left diagram provides contribution to the renormalization of $a_s$ and the right one does not yield.

Using the Ward identity and Eq.(\ref{epsilon}), one can obtain:
\begin{equation}
\Pi(L, a_s)=Z(a_s)+\Pi_B(L, a_{sB}), 
\end{equation}
where the bare polarization function reads
\begin{equation}
\label{PiB}
\Pi_B(L, a_{sB})=\sum\limits_{p\geq 1}\bigg(\frac{\mu^2}{Q^2}\bigg)^{\varepsilon p}a^{p-1}_{sB}\sum\limits_{k=-p}^{\infty}\Pi_{p,k}\varepsilon^k.
\end{equation}

Substituting (\ref{asB}), (\ref{PiB}) and expression for $Z(a_s)$ (\ref{epsilon}) in definition (\ref{RtoDint-rel}), one can arrive to the relations for the coefficients of the NS Adler function \cite{Chetyrkin:1980sa, Chetyrkin:1980pr, Gorishnii:1990vf}:
\begin{subequations}
\begin{eqnarray}
\label{d1Z}
d_1&=&-2Z_{2,-1}, \\
\label{d2Z}
d_2&=&-3Z_{3,-1}+\beta_0\Pi_{2,0}, \\
\label{d3Z}
d_3&=&-4Z_{4,-1}+2\beta_0\Pi_{3,0}+\beta_1\Pi_{2,0}+2\beta^2_0\Pi_{2,1}.
\end{eqnarray}
\end{subequations}

Comparing Eq.(\ref{D-anomalous}) with Eqs.(\ref{d1Z}-\ref{d3Z}), we conclude that the coefficients of the non-singlet contribution to the photon anomalous dimension $\gamma(a_s)$ are expressed through the first pole term $Z_{-1}(a_s)$ of the renormalization constant $Z(a_s)$ \cite{Kataev:2016aib}:
\begin{gather}
\gamma^{(M)}_{NS}(a_s)=1-\sum\limits_{p= 1}^{M} (p+1)Z_{p+1,\;-1}a^{p}_s=1+\sum\limits_{p=1}^M \gamma_p a^p_s,  \\
\text{where}~~~ \gamma^{(M)}(a_s)=d_R\bigg(\sum\limits_f Q^2_f\bigg)\gamma^{(M)}_{NS}(a_s)+d_R\bigg(\sum\limits_f Q_f\bigg)^2\gamma^{(M\geq 3)}_{SI}(a_s).
\end{gather}

Thus, one gets $\gamma_1=-2Z_{2,-1}$, $\gamma_2=-3Z_{3,-1}$, $\gamma_3=-4Z_{4,-1}$ and etc. This means that not only $\Pi(a_s)$ but also the function $\gamma(a_s)$, included  in Eq.(\ref{D-anomalous}), is related to the renormalization of the QCD charge. This fact contradicts the consideration described in \cite{Brodsky:2013vpa, Wu:2019mky}. For example, the first pole coefficient $Z_{3,-1}=C^2_F/32-133C_FC_A/576+11C_FT_Fn_f/144$ \cite{Chetyrkin:1980sa, Chetyrkin:1980pr} contains $n_f$-dependent term coming from diagrams similar to the left one presented in Figure \ref{diagrams}. The internal fermion loop, inserted into the gluon propagator,  provides obviously contributions to the renormalization of $a_s$ within the Naive-Nonabelianization procedure. In this regard, we can not treat the photon anomalous dimension $\gamma(a_s)$ as the conformal contribution containing in coefficients of the $D$-function. In the $\{\beta\}$-expansion formalism this implies that in full analogy with the coefficients of $D_{NS}$-function, the terms $\gamma_n$ of $\gamma(a_s)$ should be decomposed in coefficients of $\beta$-function  (see Eqs.(\ref{d1}-\ref{d5})):
\begin{subequations}
\begin{gather}
\label{g1}
\gamma_1 = \gamma_1[0], \\ 
\label{g2}
\gamma_2 = \beta_0 \gamma_2[1] + \gamma_2[0],  \\ 
\label{g3}
\gamma_3 = \beta_0^2 \gamma_3[2] + \beta_1 \gamma_3[0,1] + \beta_0 \gamma_3[1] + \gamma_3[0],  \\ 
\label{g4}
\gamma_4 = \beta_0^3 \gamma_4[3] + \beta_1 \beta_0 \gamma_4[1,1] + \beta_2 \gamma_4[0,0,1] + \beta_0^2 \gamma_4[2] + \beta_1  \gamma_4[0,1] + \beta_0 \gamma_4[1] + \gamma_4[0], ~ \dots
\end{gather}
\end{subequations}

This conclusion has already been made in the work \cite{Kataev:2016aib}. In accordance with this inference the values of the conformally-invariant terms $d_2[0]$, $d_3[0]$, $d_4[0]$, presented at the beginning of this section, are in full agreement with  the results of the correct $\{\beta\}$-expansion of the $D_{NS}$-function written in terms of $\gamma(a_s)$. This fact was also properly noticed in Appendix A of Ref.\cite{Brodsky:2013vpa}, but not taken into account, for instance, in the main part of \cite{Brodsky:2013vpa} and in the subsequent works \cite{Wu:2019mky, Huang:2020skl}. 

Let us focus on this issue in more detail. If one follows the logic of paper \cite{Huang:2020skl} then it appears from its results for the vector channel of the hadronic $Z$-boson decay width (see Eqs.(19), (21), (24) in this quoted paper) that contributions to the NS Adler function, called by authors of \cite{Huang:2020skl} as the ``conformally-invariant'' ones,  will be equal to:
\begin{subequations}
\begin{eqnarray}
\label{d2hat}
\hat{d}_2[0]=\gamma_2&=&-\frac{3}{32}C^2_F+\frac{133}{192}C_FC_A-\frac{11}{48}C_FT_Fn_f, \\
\label{d3hat}
\hat{d}_3[0]=\gamma_3&=&-\frac{69}{128}C^3_F+\bigg(\frac{215}{288}-\frac{11}{24}\zeta_3\bigg)C^2_FC_A+\bigg(\frac{5815}{20736}+\frac{11}{24}\zeta_3\bigg)C_FC^2_A \\ \nonumber
&+&\bigg(-\frac{169}{288}+\frac{11}{12}\zeta_3\bigg)C^2_FT_Fn_f+\bigg(-\frac{769}{5184}-\frac{11}{12}\zeta_3\bigg)C_FC_AT_Fn_f-\frac{77}{1296}C_FT^2_Fn^2_f, \\ 
\label{d4hat}
\hat{d}_4[0]=\gamma_4.
\end{eqnarray}
\end{subequations}

Comparing these results with the analogous ones, presented in Table \ref{T-d1-3}, we conclude that only the leading $C_F$-contributions to (\ref{d2hat}-\ref{d4hat}) coincide with those given in Table \ref{T-d1-3}. This is related to the fact that these contributions do not enter into expressions for the coefficients of the RG $\beta$-function. Moreover, starting from the two-loop approximation the ``conformally invariant'' terms $\hat{d}_M[0]$ in \cite{Huang:2020skl} contain the $n_f$-dependent contributions, which, as clarified by us, should be absorbed into coefficients of the RG $\beta$-function. 

This observation is confirmed by the transition to the QED limit. Indeed, in the case of the $U(1)$ gauge group with $N$ charged leptons the expressions (\ref{d2hat}-\ref{d3hat}), following from \cite{Huang:2020skl}, will turn into
\begin{subequations}
\begin{eqnarray}
\label{d2hatQED}
\hat{d}^{\;{\rm{QED}}}_2[0]&=&-\frac{3}{32}-\frac{11}{48}N, \\
\label{d3hatQED}
\hat{d}^{\;{\rm{QED}}}_3[0]&=&-\frac{69}{128}+\bigg(-\frac{169}{288}+\frac{11}{12}\zeta_3\bigg)N-\frac{77}{1296}N^2,
\end{eqnarray}
\end{subequations}
in contrast to ours:
\begin{subequations}
\begin{eqnarray}
\label{d2QED}
d^{\;{\rm{QED}}}_2[0]&=&-\frac{3}{32}, \\
\label{d3QED}
d^{\;{\rm{QED}}}_3[0]&=&-\frac{69}{128}.
\end{eqnarray}
\end{subequations}

In formulas (\ref{d2hatQED}-\ref{d3hatQED}) the fictitious ``conformally-invariant'' terms $\hat{d}^{\;{\rm{QED}}}_2[0]$, $\hat{d}^{\;{\rm{QED}}}_3[0]$, following from the results of \cite{Huang:2020skl},
contain the non-conformal $N$-dependent contributions, originating from the non-$\{\beta\}$-expanded coefficients of  $\gamma_2$ and $\gamma_3$ in QED, related to the renormalization of charge. 

Moreover, the expressions (\ref{d2hatQED}-\ref{d3hatQED}) do not correspond to the Rosner's known result \cite{Rosner:1966zz} of calculating of the divergent part of the photon field renormalization constant $Z_{ph}$ in the quenched QED, which do not contain the internal subgraphs renormalizing electromagnetic charge. The result of this work is
\begin{equation}
\bigg(Z^{-1}_{ph}\bigg)_{div}=\frac{\alpha_B}{3\pi}\bigg(1+\uuline{\frac{3}{4}}\frac{\alpha_B}{\pi}~\uuline{-\frac{3}{32}}\bigg(\frac{\alpha_B}{\pi}\bigg)^2\bigg)\ln\frac{M^2}{m^2},
\end{equation}
where $\alpha_B$ is the bare fine-structure constant, $m$ is the lepton mass and $M$ is the large scale cutoff mass. The double underlined terms are in full agreement with those, presented in Table \ref{T-d1-3} and (\ref{d2QED}), but the second underlined one contradicts the expression (\ref{d2hatQED}),  which follows from Ref.\cite{Huang:2020skl} and the related papers of these team.

Let us give one more argument in the favor of  validity of the approach requiring the decomposition of all coefficients of the photon anomalous dimension $\gamma(a_s)$ in powers of coefficients of $\beta$-function  in compliance with the $\{\beta\}$-expansion (\ref{g1}-\ref{g4}). It ensues from Eq.(\ref{D-anomalous}) that at the four-loop level the NS contributions to the Adler function and $\gamma(a_s)$ are related by the equality:
\begin{equation}
\label{d4-g4}
d_4=\gamma_4+3\beta_0\Pi_3+2\beta_1\Pi_2+\beta_2\Pi_1,
\end{equation} 
where coefficients $\Pi_n$ are defined as $\Pi_{NS}(a_s)=\sum\limits_{n\geq 0}\Pi_n a^n_s$. The terms $\gamma_4$ and $\Pi_3$ follow from the results  \cite{Baikov:2012zm}. It is interesting that their explicit expressions contain the Riemann $\zeta_4$-contributions, which, however, 
 are mutually canceled out in $d_4$ (see  results of explicit calculations of $d_4$ in \cite{Baikov:2010je}):
\begin{equation}
d^{(\zeta_4)}_4=\gamma^{(\zeta_4)}_4+3\beta_0\Pi^{(\zeta_4)}_3\equiv 0.
\end{equation}
If we properly expand $\gamma_4$ (accordingly to (\ref{g4})) and $\Pi_3$ as $\Pi_3=\beta^2_0\Pi_{3,\beta^2_0}+\beta_1\Pi_{3,\beta_1}+\beta_0\Pi_{3,\beta_0}+\Pi_{3,\beta^0_0}$,  we will naturally arrive to the absence of the $\zeta_4$-contributions in expression for $d_4[0]$ (see Table \ref{T-d4}). However, as follows from the relation (\ref{d4hat}) of Ref.\cite{Huang:2020skl} the $\hat{d}_4[0]$-term will contain $\zeta_4$-contributions\footnote{This fact immediately follows from the analytic form of $\gamma_4$
\cite{Baikov:2012zm}.}. This circumstance contradicts our outcomes and results of \cite{Cvetic:2016rot}. Moreover, even its QED counterpart $\hat{d}^{\;{\rm{QED}}}_4[0]$  \cite{Huang:2020skl} will incorporate the contribution $11\zeta_4 N^2/32$ proportional not only to $\zeta_4$-term but also to $N^2$-factor. As we have already explained above, the total $N^2$-dependence should be contained in the coefficients of the RG $\beta$-function. All these facts point to the necessity of $\{\beta\}$-decomposing all coefficients of the photon anomalous dimension if we aim to extract the conformally-invariant part of the NS Adler function in the correct way\footnote{The results \cite{Aleshin:2019yqj} of the three-loop calculations of the Adler $D$-function in
terms of the anomalous dimension of matter superfields
in the $\mathcal{N}=1$ SUSY QCD \cite{Shifman:2014cya}
 provide the extra support of this statement.}.

Another approach, leading to the results which differ from those presented in this work, consists in adding in the theory of strong interactions the extra hypothetical degrees of freedom in the form of the Majorana multiplet of light gluinos \cite{Mikhailov:2004iq, Kataev:2014jba, Mikhailov:2016feh}. Such a trick was invented in \cite{Mikhailov:2004iq} to unambiguously divide the flavor dependence in the three-loop coefficient $d_3$ between  $\beta_0$ and $\beta_1$. For this goal the expansion of the coefficient $d_3(n_f, n_{\tilde{g}})$, calculated in \cite{Chetyrkin:1996ez} for this minimally extended SUSY QCD with the number $n_{\tilde{g}}$ of light gluinos, in the $\MSbar$-scheme terms $\beta_0(n_f, n_{\tilde{g}})$ and $\beta_1(n_f, n_{\tilde{g}})$ \cite{Jones:1974pg} was considered in \cite{Mikhailov:2004iq}. This was done in close in time era, when the possibility of the existence of a light gluino has not yet been experimentally excluded. Since coefficients of the function $\beta(n_f, n_{\tilde{g}})$ include the extra degrees of freedom $n_{\tilde{g}}$  (see work \cite{Jones:1974pg} from the results of which the below expressions follow)
 \begin{subequations}
\begin{eqnarray}
\label{b0}
\beta_0(n_f, n_{\tilde{g}})&=&\frac{11}{12}C_A-\frac{1}{3}\bigg(T_Fn_f+\frac{1}{2}C_An_{\tilde{g}}\bigg)~, \\
\label{b1}
\beta_1(n_f, n_{\tilde{g}})&=&\frac{17}{24}C^2_A-\frac{5}{12}C_A\bigg(T_Fn_f+\frac{1}{2}C_An_{\tilde{g}}\bigg)-\frac{1}{4}\bigg(C_FT_Fn_f+\frac{1}{2}C^2_An_{\tilde{g}}\bigg)~,
\end{eqnarray}
\end{subequations}
then this split in such model without squarks can be performed unequivocally. However, starting from the three-loop level the $\{\beta\}$-expanded coefficients of the NS Adler function, obtained in this way, are distinct from those presented in Tables \ref{T-d1-3} and \ref{T-d4}. Indeed, the results of \cite{Kataev:2014jba} read:
\begin{subequations}
\begin{eqnarray}
\label{d10-gl}
d^{n_{\tilde{g}}}_1[0]&=&\frac{3}{4}C_F, ~~ d^{n_{\tilde{g}}}_2[1]=\bigg(\frac{33}{8}-3\zeta_3\bigg)C_F, ~~ d^{n_{\tilde{g}}}_2[0]=-\frac{3}{32}C^2_F+\frac{1}{16}C_FC_A, \\
\label{d32-gl}
d^{n_{\tilde{g}}}_3[2]&=&\bigg(\frac{151}{6}-19\zeta_3\bigg)C_F, ~~~
d^{n_{\tilde{g}}}_3[0,1]=\bigg(\frac{101}{16}-6\zeta_3\bigg)C_F, \\
\label{d31-gl}
d^{n_{\tilde{g}}}_3[1]&=&\bigg(-\frac{27}{8}-\frac{39}{4}\zeta_3+15\zeta_5\bigg)C^2_F+\bigg(-\frac{9}{64}+5\zeta_3-\frac{5}{2}\zeta_5\bigg)C_FC_A, \\
\label{d30-gl}
d^{n_{\tilde{g}}}_3[0]&=&-\frac{69}{128}C^3_F+\frac{71}{64}C^2_FC_A+\bigg(\frac{523}{768}-\frac{27}{8}\zeta_3\bigg)C_FC^2_A .
\end{eqnarray}
\end{subequations}

Taking into account the explicit expressions for terms $d_3[i]$ (see Table \ref{T-d1-3}) and for $d^{n_{\tilde{g}}}_3[i]$ (\ref{d32-gl}-\ref{d30-gl}), we can fix the difference $\Delta_3[i]$ between them in the $\MSbar$-scheme:
\begin{subequations}
\begin{eqnarray}
\Delta_{3} [i]&=& d_3[i]-d^{n_{\tilde{g}}}_3[i], \\
\label{delta_d2}
\Delta_{3} [2] &=& 0, \\
\label{delta_d0}
\Delta_{3} [0] &=&\frac{C_A}{256} 
\bigg(11C_F+7C_A\bigg)\underline{C_F(48\zeta_3-35)},\\
\label{delta_d1}
\Delta_{3} [1] &=&-\frac{1}{64} 
\bigg(3C_F+5C_A\bigg) \underline{C_F( 48\zeta_3-35)}, \\
\label{delta_d01}
\Delta_{3} [0,1] &=&\frac{1}{16}\; \underline{C_F(48\zeta_3-35)}, 
\end{eqnarray}
\end{subequations}
which turns out to be proportional to the factor $C_F(48\zeta_3-35)$. The similar analysis can be performed for the Bjorken polarized sum rule as well.

Note one interesting fact. If we fix the number  of quark flavors in accordance with the formal solution of the equation $\beta_0(n^*_f)=0$, which corresponds to the Banks-Zaks ansatz \cite{Banks:1981nn} and to the case of the effective conformal limit at one-loop level, then we will obtain  $T_Fn^*_f=11C_A/4$. In the same way, solving the equation $\beta_1(n^{**}_f)=0$, one can get $T_Fn^{**}_f=17C^2_A/(10C_A+6C_F)$. It turns out that the differences $\Delta_3[i]$ can be rewritten in the following form:
\begin{equation}
\label{n^*_f}
\Delta_3[0]=-\beta_1(n^*_f)\Delta_3[0,1], ~~~~~ \Delta_3[1]=\frac{\beta_1(n^*_f)}{\beta_0(n^{**}_f)}\Delta_3[0,1].
\end{equation} 

The similar analysis was previously performed for the quantity $d_4(n_f)+c_4(n_f)$ in Refs.\cite{Kataev:2010du, Kataev:2010tps}. The values of $d_4(n^*_f)+c_4(n^*_f)$ and $d_4(n^{**}_f)+c_4(n^{**}_f)$, obtained there with help of the $\beta$-expansion and the two-fold generalization of the Crewther relation, are in agreement with the results of the direct calculations conducted in \cite{Baikov:2010je}.

In the case of the $SU(3)$ color group the expressions (\ref{delta_d0}-\ref{delta_d01}) acquire the following form:
\begin{equation}
\label{delta-numerical}
\Delta_{3} [0]=12.6498, ~~~\Delta_{3} [1]=-8.9849, ~~~\Delta_{3} [0,1]=1.8916.
\end{equation}

The deviation $\Delta_{3} [0]$ in Eq.(\ref{delta-numerical}), obtained for the conformally invariant terms in the 3-rd order of PT, is smaller in modulus than
the value $d_3[0]=-23.2227$ from Eq.(\ref{DNS-beta-concrete}). The analogous conformal-invariant term in the theory with light gluino is $d^{n_{\tilde{g}}}_3[0]=-35.8725$ \cite{Mikhailov:2004iq, Kataev:2014jba}. The qualitative agreement between $d_3[0]$ and $d^{n_{\tilde{g}}}_3[0]$ is retained.

In its turn, the values $d_3[1]=4.9402$, $d_3[0,1]=0.6918$ from Eq.(\ref{DNS-beta-concrete}) 
differ considerably from their counterparts $d^{n_{\tilde{g}}}_3[1]=13.9251$ and $d^{n_{\tilde{g}}}_3[0,1]=-1.1997$. This fact will lead to a noticeable shift of the coupling dependent scale, defined at the stage of application of the PMC/BLM procedure.

However, nowadays, it is known that the hypothetical models with light gluinos are closed. Indeed, the comparison of the three-loop results for the hadronic $\tau$ decay and hadronic cross sections in $e^+e^-$ annihilation between $5\;{\rm{GeV}}$ and $M_Z$ with experimental data of the LEP Collaboration (CERN)
\cite{Csikor:1996vz} has indicated that the light gluinos are absent. Moreover, the analysis of the LHC data on $pp$-collisions at $\sqrt{s}=13\;{\rm{TeV}}$ (CMS Collaboration) demonstrates that gluino with mass up to $2\;{\rm{TeV}}$ is excluded at 95\% confidence level (see \cite{CMS:2019zmd, Sarkar:2021lju}). Nevertheless, such  a trick with introduction in the theory of additional degrees of freedom is very useful from perspective of super-high energy studies of  special renormalization features of
the SUSY extensions of QCD \cite{Chetyrkin:1996ez}.

The two-fold formalism does not require  the introduction of additional degrees of freedom and leads to results consistent with other independent  studies of the structure of the perturbative series for the Adler function \cite{Grunberg:1991ac, Brodsky:2011ta} and Bjorken polarized sum rule \cite{Kataev:1992jm}. 

To specify the ambiguities of application of the PMC method to the $e^+e^-$ annihilation $R$-ratio and to the Bjorken polarized sum rule in QCD,  it is highly desirable to analyze the existing data at the four-loop level taking into account the modifications of the PMC expressions described above.

\section{Conclusion}

The application of the two-fold representation of the perturbative expressions for the non-singlet Adler function and Bjorken polarized sum rule, reproducing the structure of $\{\beta\}$-decomposition, has enabled us to define 8 out of 12 possible $\{\beta\}$-expanded terms of these physical quantities in the 5-th order of PT. We demonstrate that the $\beta^3_0$, $\beta^2_0$, $\beta_0$-dependent terms and the $\beta$-independent conformally-invariant contribution remain unknown only. Up to the four-loop level results of this approach are in agreement with those, obtained with help of other methods and presented previously in literature. We emphasize that the correct $\{\beta\}$-expansion procedure requires the decomposition of the photon anomalous dimension into the coefficients of the RG $\beta$-function. Convincing justifications in the favor of this statement are given. As known, the 5-th order PT expressions for the non-singlet Adler function and Bjorken polarized sum rule contain the Riemann $\zeta_4$-contributions. We fix them in the case of the generic simple gauge group. Moreover, we
demonstrate that these $\zeta_4$-functions will be included in only three of the remaining four unknown terms of the $\{\beta\}$-expansion, namely in $\beta^2_0$, $\beta_0$-dependent and $\beta$-free coefficients. We also define their analytical Lie group structure. The arguments in the favor of  validity of these values, coming from the definite cancellations due to the scale and conformal symmetries, are given.  The results and outcomes, presented in this work, may be useful in a possible future more detailed analysis of the analytical structure of the five-loop corrections to the considered renorm-group quantities.

\section*{Acknowledgments}

We would like to thank S.V. Mikhailov for numerous useful discussions and K.V. Stepanyantz for the interest in this work. 
The work of VSM was supported by the Russian Science Foundation, agreement no. 21-71-30003.

\begin{flushleft}

\end{flushleft}
\end{document}